\documentclass[aps,prd,reprint,amsmath,amssymb]{revtex4-2}
\usepackage{mathtools,graphicx,tikz,multirow,pgfplots}
\usetikzlibrary{fit,shapes.geometric,shapes.misc}
\pgfplotsset{compat=1.3}
\usepackage{newtxtext,newtxmath}
\definecolor{citecolor}{RGB}{45,47,146}
\usepackage[colorlinks,citecolor=citecolor,anchorcolor=red,menucolor=red, linkcolor=citecolor,filecolor=red,runcolor=red,urlcolor=citecolor,frenchlinks=red]{hyperref}

\begin{document}
	\let\oldcite=\cite
	\renewcommand{\cite}[1]{\textcolor{citecolor}{\oldcite{#1}}}
	\renewcommand{\eqref}[1]{\textcolor{citecolor}{(\ref{#1})}}
	
	\title{Three-body resonances of $\alpha\alpha M$ clusters ($M=\phi$, $J/\psi$, $\eta_c$) in $\prescript{8}{M}{\mathrm{Be}}$ nuclei}
	\author{Hao Zhou$^{1,2,3}$}
	\author{Xiang Liu$^{1,2,3,4}$}\email{xiangliu@lzu.edu.cn}
	\affiliation{
		$^1$School of Physical Science and Technology, Lanzhou University, Lanzhou 730000, China\\
		$^2$Lanzhou Center for Theoretical Physics,
		Key Laboratory of Theoretical Physics of Gansu Province,
		Key Laboratory of Quantum Theory and Applications of MoE,
		Gansu Provincial Research Center for Basic Disciplines of Quantum Physics, Lanzhou University, Lanzhou 730000, China\\
		$^3$Research Center for Hadron and CSR Physics, Lanzhou University $\&$ Institute of Modern Physics of CAS, Lanzhou 730000, China\\
		$^4$MoE Frontiers Science Center for Rare Isotopes, Lanzhou University, Lanzhou 730000, China}
	
	\begin{abstract}
Motivated by the recently obtained HAL QCD potentials for the $N$-$\phi$, $N$-$J/\psi$, and $N$-$\eta_c$ interactions, we investigate the structure of the exotic nuclei $\prescript{8}{\phi}{\text{Be}}$, $\prescript{8}{J/\psi}{\text{Be}}$, and $\prescript{8}{\eta_c}{\text{Be}}$\footnote{While the notation $\prescript{9}{M}{\mathrm{Be}}$ appears in prior literature~\cite{Filikhin:2024avj, Filikhin:2024xkb, Etminan:2025yoq}, the standard convention for such symbols is that the upper-left superscript corresponds to the baryon number. Consequently, we employ the notation $\prescript{8}{M}{\mathrm{Be}}$ throughout this paper.} as $\alpha+\alpha+M$ three-body systems ($M$ denotes the meson). The bound and resonant states are calculated consistently using the Gaussian expansion method, with resonances identified via the complex scaling method. For the $\alpha\phi$ and $\alpha$-charmonium interactions, a folding potential is constructed based on the HAL QCD potentials and fitted to a Woods–Saxon form. We find that the $\phi$ meson exhibits a strong ``glue-like" effect, binding the $0^+_1$, $2^+_1$, and $4^+_1$ resonant states of $^8$Be into stable states and significantly reducing the $\alpha$-$\alpha$ distance. In contrast, the interactions of $J/\psi$ and $\eta_c$ with the nucleus are weaker, forming only shallow bound states with the $0^+_1$ state of $^8$Be and even increasing the $\alpha$-$\alpha$ separation. Notably, our analysis predicts weakly bound $\alpha$-$J/\psi$ states in the $^4S_{3/2}$ and $^2S_{1/2}$ channels, a result not reported in prior studies, which suggests that $\prescript{8}{J/\psi}{\text{Be}}$ may not be a Borromean nucleus. The sensitivity of the $\prescript{8}{\phi}{\text{Be}}(4^+_1)$ state—transitioning from bound to resonant depending on the $\alpha$-particle radius—highlights the subtle dynamics at play. These results provide a systematic theoretical comparison of how different vector mesons modify nuclear clustering, offering critical predictions for future experimental searches of such exotic hadron-nucleus systems.
	\end{abstract}
	
	\maketitle
	
	\section{INTRODUCTION}
 
	In recent decades, meson–nucleus interactions have been extensively studied both theoretically and experimentally~\cite{Hayano:2008vn,Metag:2017yuh,Krein:2017usp,Kezerashvili:2021ren}. However, the low-energy interactions of the $\phi$ meson with nucleons remain poorly understood due to the non-perturbative nature of the strong interaction and a lack of  experimental data. A recent measurement of the $N$-$\phi$ correlation function by the ALICE Collaboration revealed a strong attractive interaction between $\phi$ mesons and nucleons~\cite{ALICE:2021cpv}, sparking significant interest in this area. Furthermore, the Okubo-Zweig-Iizuka (OZI) rule suppresses interactions between charmonia and nucleons via light-meson exchange. If charmonia bind to nuclei, non-OZI mechanisms such as gluon exchange may provide the necessary attraction~\cite{Brodsky:1989jd,Wasson:1991fb}.

Advances in the HAL QCD method have enabled the extraction of interactions between nucleons and the $\phi$, $J/\psi$, and $\eta_c$ mesons from $(2+1)$-flavor lattice QCD simulations with nearly physical pion masses~\cite{Lyu:2022imf,Lyu:2024ttm}. These potentials exhibit a universally attractive character across all distance scales, from short-range cores to long-range tails, offering crucial insights into non-perturbative QCD dynamics. Based on these HAL QCD potentials, we can now investigate the energy spectrum structures and dynamical features of meson–nucleus systems in greater detail.

The study of nuclear-bound quarkonium dates back to 1990. Ref.~\cite{Brodsky:1989jd} suggested that charmonium binds to nuclei with mass number $A\ge 3$, with a binding energy increasing with $A$ and reaching about 400 MeV for $A=9$. In contrast, Ref.~\cite{Wasson:1991fb} argued that the binding energy does not increase strongly with $A$, instead saturating at a maximum of $\sim30$ MeV. Recently, using the HAL QCD potential, calculations of $\phi NN$ systems via the complex scaling method~\cite{Wen:2025wit} and Faddeev equations~\cite{Filikhin:2024avj} indicate that $\phi$ can form deep bound states with one or two nucleons. Ref.~\cite{Wen:2025wit} also studied $J/\psi NN$ and $\eta_cNN$ systems, estimating that charmonium can form bound states with approximately 3–4 nucleons. Subsequently, bound states of larger meson–nucleus systems, namely $\prescript{8}{\phi}{\text{Be}}$, $\prescript{8}{J/\psi}{\text{Be}}$, and $\prescript{8}{\eta_c}{\text{Be}}$, were calculated in Refs.~\cite{Filikhin:2024xkb,Etminan:2025yoq}. However, the resonant states of these systems and the dynamic effects imparted by the mesons remain unexplored in the literature.

A primary goal of hypernuclear physics is to extract information on hadron-hadron interactions~\cite{Hiyama:1997ub,Hiyama:2009zz,Hiyama:2012sq}, close to our understanding of nonperturbative strong interaction. Once an interaction potential is determined, the energy spectrum of the system can be calculated. By comparing theoretical spectra with experimental data, one can infer details about the underlying interactions. For instance, Ref.~\cite{Hiyama:2000jd} successfully inferred the $\Lambda N$ spin-orbit force from the energy spectra of $\prescript{9}{\Lambda}{\text{Be}}$ and $\prescript{13}{\Lambda}{\text{C}}$. A similar approach can be applied to the $\prescript{8}{\phi}{\text{Be}}$, $\prescript{8}{J/\psi}{\text{Be}}$, and $\prescript{8}{\eta_c}{\text{Be}}$ systems. Moreover, experimental measurements of spectra for these systems would provide valuable validation for the HAL QCD potentials.

Another key objective is to study new dynamical features induced by the introduction of a non-nucleonic particle~\cite{Hiyama:1996gv}. Since the Pauli exclusion principle does not operate between non-nucleonic particles and nucleons, their inclusion can lead to additional bound states and a significant contraction of the nuclear core—a phenomenon we term the ``glue-like" role of the particle. Previous work has largely focused on the $\Lambda$ hyperon~\cite{Motoba:1983kbv,Motoba:1985,Bando:1983pv}, but particles containing strange or even heavier charm quarks may also exhibit such glue-like effects. Therefore, we will explore the glue-like roles of $\phi$, $J/\psi$, and $\eta_c$ particles within $\prescript{8}{M}{\mathrm{Be}}\ (M=\phi$, $J/\psi$, $\eta_c)$ nuclei and compare them with the $\Lambda$ hyperon.

Directly solving the Schr\"odinger equation for systems with particle numbers $n>5$ is computationally intractable. Consequently, we must employ approximation methods. The cluster model, which considers a nuclear system as a combination of several subunits (clusters), allows its structure to be studied using realistic interactions and advanced computational techniques. The recent experimental validation of the $\prescript{10}{}{\text{Be}}$ ground-state molecular structure~\cite{Li:2023msp} further strengthens confidence in such cluster approaches. In this picture, $^8$Be is considered as two $\alpha$ clusters, while $\prescript{8}{M}{\mathrm{Be}}$ is treated as an $\alpha\alpha M$ three-cluster system. This framework allows us to analyze these nuclei using well-developed methods for two- and three-body systems.

The Gaussian expansion method (GEM)~\cite{Hiyama:2003cu,Hiyama:2012sma} is a high-precision variational technique for solving the Schr\"odinger equation of few-body systems. The GEM achieves numerical precision comparable to the Faddeev-Yakubovsky method for systems like $^3$H ($^3$He) and $^4$He, and has successfully described diverse atomic, baryonic, and quark few-body systems. To theoretically obtain resonant states, several approaches have been developed, among which the complex scaling method (CSM)~\cite{Aguilar:1971ve,Balslev:1971vb,Simon:1972} has been particularly successful. The CSM is widely regarded as one of the few, if not the only, practical methods for treating many-body resonances beyond two-particle systems~\cite{Aoyama:2006hrz,Myo:2014ypa}. By applying the CSM and solving the resulting complex-scaled Schr\"odinger equation with the Gaussian expansion method, one can obtain the resonant states of the $\prescript{8}{M}{\mathrm{Be}}$ system.

This paper is organized as follows. Following the introduction, Sec. \ref{sec2} discusses the treatment of three-body resonances in $\alpha\alpha M$ clusters within $\prescript{8}{M}{\mathrm{Be}}$. Next, Sec. \ref{sec3} presents the numerical results. Finally, the paper concludes with a summary.
    
	\section{Three-body resonances of $\alpha\alpha M$ cluster}\label{sec2}
	
	\subsection{Hamiltonian}
	In the framework of the $\alpha+\alpha+M$ three-body cluster model, we need to know the interaction between $\alpha$ and $\alpha$, as well as the interaction between $\alpha$ and $M$. 
	
	\textit{$\alpha$-$\alpha$ interaction.} There are usually two methods to describe the $\alpha$-$\alpha$ interaction: the orthogonality condition model (OCM)~\cite{Saito:1969zz} and the phenomenological potential model. The orthogonality condition model folds the effective $NN$ potential and $pp$ Coulomb potential into the $\alpha$ cluster wave function and takes the Pauli principle between two $\alpha$ clusters into account through the OCM projection operator~\cite{Lee:2019mlt,Wu:2019ivs}. The phenomenological potential model divides the $\alpha$-$\alpha$ interaction into the nuclear and Coulomb parts:
	\begin{equation}
		V(r) = V_N(r) + V_C(r).
	\end{equation}
	Different potential models adopt different functional forms for these two parts~\cite{Awasthi:2023yre}. For the nuclear part, there are the Morse potential~\cite{Morse:1929zz}, the Woods-Saxon potential~\cite{Darriulat:1965zz}, the double Gaussian Ali-Bodmer potential~\cite{Ali:1966olw}, the Malfliet-Tjon potential~\cite{Malfliet:1968tj}, the double Hulthén potential~\cite{Bhoi:2016wge}, and the double exponential~\cite{Awasthi:2023yre}. For the Coulomb part, there are the Coulomb potential, the atomic Hulthén potential~\cite{hulthen:1942}, and the modified Coulomb potential~\cite{Herzenberg:1957}. We use the same phenomenological $\alpha$-$\alpha$ potential as in Ref.~\cite{Ali:1966olw}. Its nuclear part is double Gaussian potential: 
	\begin{equation}
		V_{N}(r) = V_r\mathrm{e}^{-\mu_r^2r^2}-V_a\mathrm{e}^{-\mu_a^2 r^2},
	\end{equation}
	where $V_r$ and $V_a$ represents the strength of repulsive and attractive parts, respectively, $\mu_r$ and $\mu_a$ are their corresponding inverse ranges. 
	Its Coulomb part is
	\begin{equation}\label{eq:Coulomb part}
		V_C(r)=\frac{4\alpha}{r},
	\end{equation}
	where $\alpha$ is the fine structure constant. 
	
	The parameters in the $\alpha$-$\alpha$ potential are shown in Table~\ref{tab:parameters}. It can be noted that the parameters we used are different from those in Refs.~\cite{Filikhin:2024xkb,Etminan:2025yoq}. They selected the parameter set (a) from Ref.~\cite{Ali:1966olw}, but this set of parameters did not fit well for $\alpha$-$\alpha$ scattering phase shift. In fact, Ref.~\cite{Ali:1966olw} recommends the parameter set (d) we are currently using. Furthermore, according to our calculations, the parameter set (a) cannot accurately reproduce the ground state of Be. Fortunately, the parameter set (a) has a minimal impact ($\sim$0.1 MeV) on the binding energy of $\prescript{8}{M}{\mathrm{Be}}$, but a significant one on its resonant states. 
	
	\textit{$N$-$\phi$, $N$-$J/\psi$, and $N$-$\eta_c$ interaction.} For the $N$-$\phi$, $N$-$J/\psi$, and $N$-$\eta_c$ interactions, we adopt the potential from the HAL QCD, which employed three Gaussian functions for fitting. The potentials are expressed as
	\begin{equation}\label{eq:threeGaussian}
		V_{NM}^J(r)=\sum_{i=1}^{3} a_i\mathrm{e}^{-\frac{r^2}{b_i^2}},
	\end{equation}
	where the symbol $M$ denotes $\phi$, $J/\psi$, or $\eta_c$. {
For $\phi$ and $J/\psi$, the superscript $J$ can take the values $1/2$, $3/2$, or $\text{ave}$; for $\eta_c$, no $J$ superscript is used.
Here $J=1/2$, $3/2$, and $\text{ave}$ correspond respectively to the ${}^2S_{1/2}$ channel, the ${}^4S_{3/2}$ channel, and the spin-averaged $N$-$M$ interaction.
The potential $V_{NM}^\text{ave}(r)$ is defined by Eq.~\eqref{eq:Vave}.} In addition to the Gaussian form of the potential mentioned above, the $N$-$\phi$ interaction in the ${}^4S_{3/2}$ channel is also fitted by a function form with the two-pion exchange tail, namely
	\begin{equation}
		V_{N\phi}^{3/2}(r)=\sum_{i=1}^{2}a_i\mathrm{e}^ {-(\frac{r}{b_i})^{2}}+a_{3}m_{\pi}^{4}f(r;b_3)\frac{\mathrm{e}^{-2m_\pi r}}{r^2}. \label{eq:VphiN3/2}
	\end{equation}
	For the $N$-$\phi$ interaction in the ${}^2S_{1/2}$ channel, its real part adds a parameter $\beta$ compared to the ${}^4S_{3/2}$ channel: 
	\begin{equation}\label{eq:Nphi1/2}
		V_{N\phi}^{1/2}(r)=\beta\sum_{i=1}^2a_i\mathrm{e}^ {-(\frac{r}{b_i})^{2}}+a_{3}m_{\pi}^{4}f(r;b_3)\frac{\mathrm{e}^{-2m_\pi r}}{r^2},
	\end{equation}
	where the parameter $\beta=6.9_{-0.5}^{+0.9}$(stat.)${}_{-0.1}^{+0.2}$(syst.) which is determined via a fit to the experimental $p$–$\phi$ correlation function~\cite{Chizzali:2022pjd}, measured by the ALICE collaboration in $pp$ collisions at $\sqrt{s}=13$ TeV~\cite{ALICE:2021cpv}, and Argonne-type form factor $f(r;b_3)=(1-\mathrm{e}^{-(\frac{r}{b_3})^2})^2$. 
	The parameters are collected in Table~\ref{tab:parameters}. 
	
	\textit{$\alpha$-$\phi$, $\alpha$-$J/\psi$, and $\alpha$-$\eta_c$ interaction.} The $\alpha$-$\phi$, $J/\psi$, and $\eta_c$ potentials are approximated by folding the $N$-$\phi$, $J/\psi$, and $\eta_c$ potential into the nucleon density function of the $\alpha$ particle as done with the single folding model~\cite{Satchler:1979ni}. {The folding procedure effectively convolves the nucleon-hadron interaction with the density distribution of the $\alpha$-particles, yielding an effective $\alpha$-hadron potential.} The $\alpha$-$M$ potential
	\begin{equation}\label{eq:alpha-qq}
		V_{\alpha M}^J(\boldsymbol{r})=\int\rho(\boldsymbol{x})V_{NM}^J(\boldsymbol{r}-\boldsymbol{x})\mathrm{d}\boldsymbol{x},
	\end{equation}
	where $\rho(\boldsymbol{x})$ is the nucleon density function of $\alpha$ particle in the center-of-mass frame, and $\boldsymbol{r}$ is the vector from the center-of-mass of the $\alpha$ particle to $M$. Following Ref.~\cite{Wang:2023uek}, we choose the simple Gaussian distribution
	\begin{equation}
		\rho(\boldsymbol{x})=4\left(\frac{1}{\pi a^2}\right)^{\frac{3}{2}}\mathrm{e}^{-\frac{x^2}{a^2}}
	\end{equation}
	as the nucleon density function, where $a=\sqrt{\frac{2}{3}}R_\alpha$ ($R_\alpha$ is the rms matter radius of $\alpha$ particles). For the three Gaussian potential~\eqref{eq:threeGaussian}, we can obtain analytical expressions for Eq.~\eqref{eq:alpha-qq}:
	\begin{equation}\label{eq:analytical alpha-qq}
		V_{\alpha M}^J(r)=\sum_{i=1}^{3}\frac{4a_ib_i^3\mathrm{e}^{-\frac{r^2}{a^2+b_i^2}}}{(a^2+b_i^2)^{3/2}}.
	\end{equation}
	
	As explained in Ref.~\cite{Filikhin:2024xkb}, there is a certain degree of error in the root-mean-square (rms) charge radius and matter radius of $\alpha$ particles in the experiment. We have three experimental values 1.681(4) fm~\cite{Sick:2008zza}, 1.6755(28) fm~\cite{Angeli:2013epw}, and 1.67824(83) fm~\cite{Krauth:2021foz} for the rms charge radius. Recently, the rms matter radius of $\alpha$ particles was measured to be 1.70 ± 0.14 fm~\cite{Wang:2023uek}. Hence, we calculated the energy spectra of the system for $R_\alpha=1.84$, 1.70, and 1.56 fm separately. 
 
	For the $N$-$\phi$ interaction in the ${}^2S_{1/2}$ channel, we only have Eq.~\eqref{eq:Nphi1/2} that cannot be analytically integrated in Eq.~\eqref{eq:alpha-qq}. Therefore, we obtain the $\alpha$-$\phi$ potential in the ${}^2S_{1/2}$ channel by fitting the numerical results of Eq.~\eqref{eq:alpha-qq} using the Woods-Saxon potential~\cite{Dover:1982ng}
	\begin{equation}\label{eq:WS potential}
		V_{\alpha M}^J(r)=-\frac{V_0}{1+\mathrm{e}^{\frac{r-R}{c}}},
	\end{equation}
	where $V_0$ is the strength of the interaction, $R$ is the radius of the nucleus, and $c$ is the surface diffuseness. The fitted parameters when $R_\alpha=1.84$, 1.70, and 1.56 fm are shown in Table~\ref{tab:parameters}. In practical calculations, we use Eq.~\eqref{eq:WS potential} for the $\alpha$-$\phi$ potential in the ${}^2S_{1/2}$ channel, and analytical expression Eq.~\eqref{eq:analytical alpha-qq} for the potentials of other systems. 

\begin{figure}[htbp]
		\begin{tikzpicture}
			\begin{axis}[
				xlabel=$r$ (fm),
				xmin=0, xmax=5, 
				ylabel=$V_{\alpha\phi}(r)$ (MeV),
				ymin=-60, 
				ytick distance=10, minor y tick num=4,
				legend entries={$N\phi({}^4S_{3/2})$,$R_\alpha=1.84$ fm, $R_\alpha=1.70$ fm, $R_\alpha=1.56$ fm},
				legend pos=south east,
				thick
				]
				\newcommand\al{-371}
				\newcommand\bl{0.15}
				\newcommand\az{-50}
				\newcommand\bz{0.66}
				\newcommand\as{-31}
				\newcommand\bs{1.09}
				\addplot+[domain=0.4:5,samples=300,mark=none,dotted,cyan]{\al*exp(-x*x/(\bl*\bl))+\az*exp(-x*x/(\bz*\bz))+\as*exp(-x*x/(\bs*\bs))};
				\newcommand\ax{1.84*sqrt(2/3)}
				\addplot+[domain=0:5,samples=300,mark=none,dashed,blue]{4*\al*pow(\bl,3)/pow(\ax*\ax+\bl*\bl,1.5)*exp(-x*x/(\ax*\ax+\bl*\bl))+4*\az*pow(\bz,3)/pow(\ax*\ax+\bz*\bz,1.5)*exp(-x*x/(\ax*\ax+\bz*\bz))+4*\as*pow(\bs,3)/pow(\ax*\ax+\bs*\bs,1.5)*exp(-x*x/(\ax*\ax+\bs*\bs))};
				\renewcommand\ax{1.70*sqrt(2/3)}
				\addplot+[domain=0:5,samples=300,mark=none,red]{4*\al*pow(\bl,3)/pow(\ax*\ax+\bl*\bl,1.5)*exp(-x*x/(\ax*\ax+\bl*\bl))+4*\az*pow(\bz,3)/pow(\ax*\ax+\bz*\bz,1.5)*exp(-x*x/(\ax*\ax+\bz*\bz))+4*\as*pow(\bs,3)/pow(\ax*\ax+\bs*\bs,1.5)*exp(-x*x/(\ax*\ax+\bs*\bs))};
				\renewcommand\ax{1.56*sqrt(2/3)}
				\addplot+[domain=0:5,samples=300,mark=none,dash dot]{4*\al*pow(\bl,3)/pow(\ax*\ax+\bl*\bl,1.5)*exp(-x*x/(\ax*\ax+\bl*\bl))+4*\az*pow(\bz,3)/pow(\ax*\ax+\bz*\bz,1.5)*exp(-x*x/(\ax*\ax+\bz*\bz))+4*\as*pow(\bs,3)/pow(\ax*\ax+\bs*\bs,1.5)*exp(-x*x/(\ax*\ax+\bs*\bs))};
				\draw[gray,thin](axis cs:0,0)--(axis cs:5,0);
			\end{axis}
		\end{tikzpicture}
        \begin{tikzpicture}
			\begin{axis}[
				xlabel=$r$ (fm),
				xmin=0, xmax=5, 
				ylabel=$V_{\alpha J/\psi}(r)$ (MeV),
				ytick distance=5, minor y tick num=4,
				legend entries={Ref.~\cite{Etminan:2025yoq},$R_\alpha=1.70\sqrt{1.5}$ fm,$R_\alpha=1.70$ fm,$R_\alpha=1.47$ fm},
				legend pos=south east,
                legend cell align=left,
				thick
				]
				\addplot+[domain=0:5,samples=300,mark=none,dotted,cyan]{-8.95/(1+exp((x-1.684)/0.367))};
				\newcommand\ax{1.70}
				\newcommand\al{-51}
				\newcommand\bl{0.09}
				\newcommand\az{-13}
				\newcommand\bz{0.49}
				\newcommand\as{-22}
				\newcommand\bs{0.82}
				\addplot+[domain=0:5,samples=300,mark=none,dashed,blue]{4*\al*pow(\bl,3)/pow(\ax*\ax+\bl*\bl,1.5)*exp(-x*x/(\ax*\ax+\bl*\bl))+4*\az*pow(\bz,3)/pow(\ax*\ax+\bz*\bz,1.5)*exp(-x*x/(\ax*\ax+\bz*\bz))+4*\as*pow(\bs,3)/pow(\ax*\ax+\bs*\bs,1.5)*exp(-x*x/(\ax*\ax+\bs*\bs))};
				\renewcommand\ax{1.70*sqrt(2/3)}
				\addplot+[domain=0:5,samples=300,mark=none,red]{4*\al*pow(\bl,3)/pow(\ax*\ax+\bl*\bl,1.5)*exp(-x*x/(\ax*\ax+\bl*\bl))+4*\az*pow(\bz,3)/pow(\ax*\ax+\bz*\bz,1.5)*exp(-x*x/(\ax*\ax+\bz*\bz))+4*\as*pow(\bs,3)/pow(\ax*\ax+\bs*\bs,1.5)*exp(-x*x/(\ax*\ax+\bs*\bs))};
				\renewcommand\ax{1.47*sqrt(2/3)}
				\addplot+[domain=0:5,samples=300,mark=none,black,dash dot]{4*\al*pow(\bl,3)/pow(\ax*\ax+\bl*\bl,1.5)*exp(-x*x/(\ax*\ax+\bl*\bl))+4*\az*pow(\bz,3)/pow(\ax*\ax+\bz*\bz,1.5)*exp(-x*x/(\ax*\ax+\bz*\bz))+4*\as*pow(\bs,3)/pow(\ax*\ax+\bs*\bs,1.5)*exp(-x*x/(\ax*\ax+\bs*\bs))};
				
				\draw[gray,thin](axis cs:0,0)--(axis cs:5,0);
			\end{axis}
		\end{tikzpicture}
		\caption{The $V_{\alpha\phi}(r)$/$V_{\alpha J/\psi}(r)$ potentials corresponding to the $N$-$\phi$/$J/\psi$ interaction in the ${}^4S_{3/2}$ channel when the $\alpha$-particle rms matter radius $R_\alpha=1.84$, 1.70, and 1.56 fm/$1.70\sqrt{1.5}$, 1.70, and 1.47 fm. \label{fig:alpha-phi}}
	\end{figure}
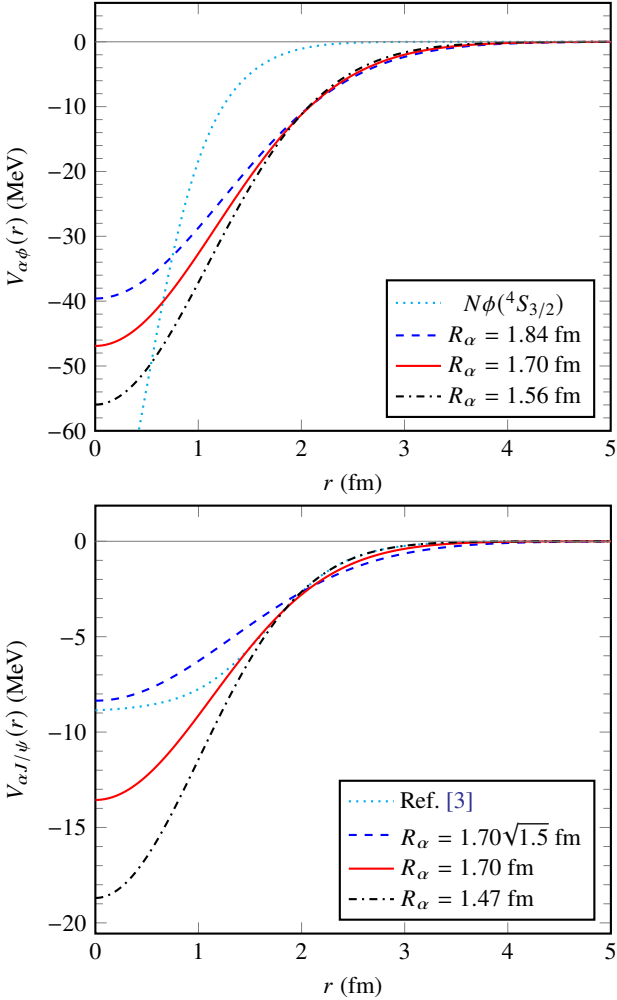

    The $N$-$\phi$ potential in the ${}^4S_{3/2}$ channel and the corresponding $\alpha$-$\phi$ potentials calculated by Eq.~\eqref{eq:analytical alpha-qq} when $R_\alpha=1.84$, 1.70, and 1.56 fm are plotted in the upper panel of Fig.~\ref{fig:alpha-phi}, which is consistent with the results obtained using Woods-Saxon fitting in Ref.~\cite{Filikhin:2024xkb}. The $V_{\alpha\phi}$ potential is highly sensitive to $R_\alpha$, and as $R_\alpha$ decreases, $V_{\alpha\phi}$ becomes deeper and deeper, which will have a significant impact on the energy spectra of the systems. The lower panel of Fig.~\ref{fig:alpha-phi} presents the $\alpha$-$J/\psi$ Woods-Saxon potential fitted by Ref.~\cite{Etminan:2025yoq} and the $\alpha$-$J/\psi$ potential calculated by Eq.~\eqref{eq:analytical alpha-qq} with $R_\alpha=1.70\sqrt{1.5}$, 1.70, and 1.47 fm in the ${}^4S_{3/2}$ channel. We choose $1.70\sqrt{1.5}$ fm because $R_\alpha$ is $\sqrt{1.5}$ times $a$, while we choose 1.47 fm because it was used to compute the rms matter radius of the $c\bar{c}$-$\alpha\alpha$ state in Ref.~\cite{Etminan:2025yoq}. It can be seen that the potential with $R_\alpha=$1.70 fm, which should have been consistent with Ref.~\cite{Etminan:2025yoq}, is deeper. When $R_\alpha=1.70\sqrt{1.5}$ fm, the two results are similar but still exhibit differences in line shape. The precise origin of these differences remains unclear; nevertheless, our results for the $V_{\alpha\phi}$ potential are consistent with Ref.~\cite{Filikhin:2024xkb}.

    {
    The earlier study of Ref.~\cite{Filikhin:2024xkb} employed a folded potential derived solely from the $^4S_{3/2}$ channel, whereas Ref.~\cite{Etminan:2025yoq} performed separate folding for the $^2S_{1/2}$, $^4S_{3/2}$, and spin-averaged channels. From a physical perspective, all channels can contribute simultaneously to the $\alpha$-$M$ interaction, and a properly spin-averaged interaction should indeed be used. To treat the spin dependence of the $N$-$M$ interaction, the standard approach is to take a spin average weighted by $2S+1$:
    \begin{equation}\label{eq:Vave}
        V_{NM}^{\text{ave}}(r)=\frac{1}{3}V_{NM}^{1/2}(r)+\frac{2}{3}V_{NM}^{3/2}(r).
    \end{equation}
    From the viewpoint of strict angular‑momentum coupling, this weighting is also well‑founded. The spin wave function of the $\alpha$ particle is
    \begin{equation}
        \chi_\alpha=[[s_1s_2]_0[s_3s_4]_0]_{00}=[s_1s_2]_{00}[s_3s_4]_{00},
    \end{equation}
    where $s_i$ denotes the spin of the $i$‑th nucleon. The total spin wave function of the $\alpha$ particle and a spin $s_5=1$ meson ($\phi$ or $J/\psi$) can be written as
    \begin{equation}
    \begin{aligned}
        \chi_{\alpha M}&=[[[s_1s_2]_0[s_3s_4]_0]_0s_5]_{1m}\\
        &=[[s_1s_2]_0s_5]_{1m}[s_3s_4]_{00}=[s_1s_2]_{00}[[s_3s_4]_0s_5]_{1m}.
    \end{aligned}
    \end{equation}
    As illustrated in Fig.~\ref{fig:Spin}, the factor $[[s_3s_4]_0 s_5]_{1m}$ can be recoupled into a superposition of $[s_3[s_4s_5]_{1/2}]_{1m}$ and $[s_3[s_4s_5]_{3/2}]_{1m}$. The resulting recoupling coefficients are exactly the weights $1/3$ and $2/3$ appearing in Eq.~\eqref{eq:Vave}, thereby confirming the consistency of the spin‑averaging prescription.
    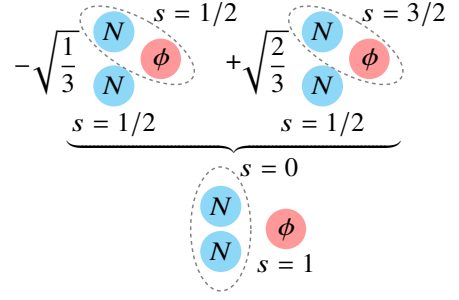
\begin{figure}[htbp]
    \resizebox{0.7\linewidth}{!}{\begin{tikzpicture}[font=\scriptsize,N/.style={circle,fill=cyan!40,inner sep=1pt},phi/.style={circle,fill=red!40,inner sep=1pt},ellip/.style={ellipse,draw=gray,line width=0.4pt,dash pattern=on 1pt off 1pt,inner sep=0pt}]
        \begin{scope}[shift={(120:2)}]
            \node(N1)at(120:0.3)[N]{$N$};
            \node(N2)at(240:0.3)[N,label={[label distance=-2pt]-90:$s=1/2$}]{$N$};
            \node(phi)at(0:0.3)[phi]{$\phi$};
            \node[ellip,rotate fit=60,fit=(N1)(phi),label={[label distance=-5pt]0:$s=1/2$}]{};
        \end{scope}
        \begin{scope}[shift={(60:2)}]
            \node(N1)at(120:0.3)[N]{$N$};
            \node(N2)at(240:0.3)[N,label={[label distance=-2pt]-90:$s=1/2$}]{$N$};
            \node(phi)at(0:0.3)[phi]{$\phi$};
            \node[ellip,rotate fit=60,fit=(N1)(phi),label={[label distance=-5pt]0:$s=3/2$}]{};
        \end{scope}
        \begin{scope}[shift={(-90:-0.1)}]
            \node(N1)at(120:0.25)[N]{$N$};
            \node(N2)at(240:0.25)[N]{$N$};
            \node(phi)at(0:0.5)[phi,label={[label distance=-2pt]-90:$s=1$}]{$\phi$};
            \node[ellip,fit=(N1)(N2),label={[label distance=-5pt]80:$s=0$}]{};
        \end{scope}
        \node at(0,0.9){$\underbrace{\hspace{3.2cm}}$};
        \node at(0,1.7){$+$};
        \node at(-1.8,1.7){$-\sqrt{\dfrac{1}{3}}$};
        \node at(0.3,1.7){$\sqrt{\dfrac{2}{3}}$};
    \end{tikzpicture}}
    \caption{Recoupling of $[[s_3s_4]_0 s_5]_{1m}$ into a superposition of $[s_3[s_4s_5]_{1/2}]_{1m}$ and $[s_3[s_4s_5]_{3/2}]_{1m}$. The coefficients $-\sqrt{1/3}$ and $\sqrt{2/3}$ yield the statistical weights $1/3$ and $2/3$ used in the spin‑averaged potential.}\label{fig:Spin}
\end{figure}
    }
	
	\begin{table*}
		\caption{The parameters of $N$-$\phi$, $N$-$J/\psi$, $N$-$\eta_c$ potential $a_1,a_2,a_3$, and $b_1,b_2,b_3$. \label{tab:parameters}}
		\begin{ruledtabular}
		\begin{tabular}{ccccccccc}
			                      \multicolumn{5}{c}{$\alpha$-$\alpha$ potential \cite{Ali:1966olw}}                       &                 \multicolumn{4}{c}{Woods-Saxon potential}                  \\ \cline{1-5}\cline{6-9}
			         $l$           & $V_r$ (MeV) & $\mu_r\ (\mathrm{fm}^{-1})$ & $V_a$ (MeV) & $\mu_a\ (\mathrm{fm}^{-1})$ &                 $R_\alpha$ (fm)                  & $V_0$ (MeV) & $R$ (fm) & $c$ (fm) \\ \cline{1-5}\cline{6-9}
			          0            &     500     &             0.7             &     130     &            0.475            &                       1.84                       &   75.9      &     1.29 &    0.559 \\
			          2            &     320     &             0.7             &     130     &            0.475            &                       1.70                       &   92.1      &    1.20   &   0.526   \\
			          4            &             &                             &     130     &            0.475            &                       1.56                       &   113      &     1.11 &    0.492 \\
			                                                       \multicolumn{9}{c}{$N$-$\phi$, $N$-$J/\psi$, $N$-$\eta_c$ potential \cite{Lyu:2022imf,Lyu:2024ttm}}                                                        \\ \hline
			                       &Interaction& $a_1$ (MeV) &         $b_1$ (fm)          & $a_2$ (MeV) &         $b_2$ (fm)          & $a_3m_\pi^{4n}\ (\text{MeV}\cdot\text{fm}^{2n})$ & $b_3$ (fm)  &                 \\ \cline{2-8}
			 &$V_{N\phi}^{3/2}(r)$  &  $-$371(19)   &           0.15(3)           &   $-$50(35)   &          0.66(61)           &                     $-$31(53)                      &  1.09(41)   &                   \\
			 &$V_{N\phi}^{1/2}(r)$  &  $-$371(27)   &           0.13(1)           &  $-$119(39)   &           0.30(5)           &                     $-$97(14)                      &   0.63(4)   &                  \\
			&$V_{NJ/\psi}^{3/2}(r)$ &   $-$51(1)    &           0.09(1)           &   $-$13(6)    &           0.49(7)           &                      $-$22(5)                      &   0.82(6)   &                   \\
			&$V_{NJ/\psi}^{1/2}(r)$ &   $-$101(1)   &           0.13(1)           &   $-$33(6)    &           0.44(5)           &                      $-$23(8)                      &   0.83(9)   &                 \\
			&$V_{N\eta_c}(r)$ &  $-$264(14)   &           0.11(1)           &   $-$28(13)   &           0.24(6)           &                      $-$22(2)                      &   0.77(3)   &                  \\ \hline
			      \multicolumn{9}{c}{$m_\alpha=3727.3794118$ MeV~\cite{Mohr:2024kco}\quad $m_\phi=1019.460$ MeV\quad $m_{J/\psi}=3096.900$ MeV\quad $m_{\eta_c}=2984.1$ MeV\quad $m_\pi=139.57039$ MeV~\cite{ParticleDataGroup:2024}}\\
		\end{tabular}
		\end{ruledtabular}
	\end{table*}
	
	\subsection{Gaussian Expansion Method}
	\begin{figure}
		\begin{tikzpicture}[semithick]
			\fill(-1,0)node[below]{$m_1$}circle[radius=1.5pt];
			\fill(1,0)node[below]{$m_2$}circle[radius=1.5pt];
			\fill(0,1.732)node[above]{$m_3$}circle[radius=1.5pt];
			\draw[-latex](-1,0)--node[above left=-1.5pt]{$\boldsymbol{r}_2$}(0,1.732);
			\draw[-latex](-0.6,0.693)--node[above]{$\boldsymbol{R}_2$}(1,0);
			\node at(0,-0.7){$c=2$};
			\begin{scope}[xshift=-3cm]
				\fill(-1,0)node[below]{$m_1$}circle[radius=1.5pt];
				\fill(1,0)node[below]{$m_2$}circle[radius=1.5pt];
				\fill(0,1.732)node[above]{$m_3$}circle[radius=1.5pt];
				\draw[-latex](0,1.732)--node[above right=-1.5pt]{$\boldsymbol{r}_1$}(1,0);
				\draw[-latex](0.6,0.693)--node[above]{$\boldsymbol{R}_1$}(-1,0);
				\node at(0,-0.7){$c=1$};
			\end{scope}
			\begin{scope}[xshift=3cm]
				\fill(-1,0)node[below]{$m_1$}circle[radius=1.5pt];
				\fill(1,0)node[below]{$m_2$}circle[radius=1.5pt];
				\fill(0,1.732)node[above]{$m_3$}circle[radius=1.5pt];
				\draw[-latex](1,0)--node[below]{$\boldsymbol{r}_3$}(-1,0);
				\draw[-latex](0,0)--node[right]{$\boldsymbol{R}_3$}(0,1.732);
				\node at(0,-0.7){$c=3$};
			\end{scope}
		\end{tikzpicture}
		\caption{Three Jacobian coordinates of three-body system.\label{fig:Jacobian coordinates}}
	\end{figure}
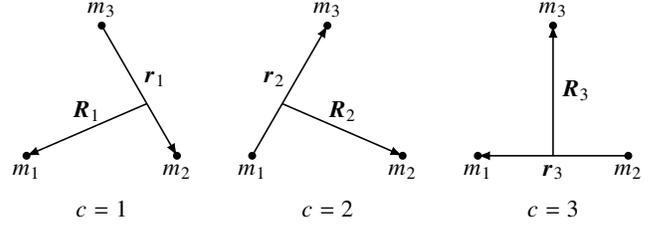
	
	A system with Hamiltonian 
$H$ is described by the stationary Schr\"odinger equation
	\begin{equation}
		H\Psi=E\Psi.
	\end{equation}
	We expand the total wave function in terms of a set of $L^2$-integrable basis functions $\Phi_\alpha$, where $\alpha$ denotes the set of quantum numbers labeling $\Phi_\alpha$, as
	\begin{equation}
		\Psi=\sum_\alpha C_\alpha\Phi_\alpha.
	\end{equation}
	The Rayleigh-Ritz variational principle then leads to a generalized matrix eigenvalue problem 
	\begin{equation}
		HC=NCE,
	\end{equation}
	where the matrix elements of the Hamiltonian and overlap are given by
	\begin{align}
		&H_{\alpha'\alpha}=\langle\Phi_{\alpha'}|H|\Phi_\alpha\rangle,\label{eq:the energy matrix element}\\
		&N_{\alpha'\alpha}=\langle\Phi_{\alpha'}|\Phi_\alpha\rangle.
	\end{align}
	Concrete calculations require the explicit choice of basis functions $\Phi_\alpha$. In the GEM, a suitably chosen set of Gaussian basis functions forms an approximately complete set within a finite coordinate space. This choice enables an accurate description of short-range correlations, long-range asymptotic behavior, and the highly oscillatory character of wave functions for both bound and scattering states of the system \cite{Hiyama:2003cu,Hiyama:2012sma}.
	
	For $\alpha+\alpha+M$ systems, we place the two alpha particles at the positions labeled $m_1$ and $m_2$ in Fig. \ref{fig:Jacobian coordinates}, respectively. Taking into account of the symmetry between the two alpha particles, we adopt the basis functions
	\begin{equation}\label{eq:Phi2}
		\Phi_\alpha=[\phi_{nl}^{\mathrm{G}}(\boldsymbol{r}_3)\phi_{NL}^{\mathrm{G}}(\boldsymbol{R}_3)]_{L_tM_t},
	\end{equation}
	with $\alpha=\{l,L,L_t,n,N\}$. The values of $\alpha$ used for each $J^P$ quantum number are listed in Table \ref{tab:Angular-momentum space}. 
    \begin{table}[htbp]
		\caption{Three-body angular-momentum configuration space of Be. The parameters $\nu_1$ and $\lambda_1$ have units of $\text{fm}^{-2}$; all other quantities are dimensionless. \label{tab:Angular-momentum space}}
		\begin{ruledtabular}
			\begin{tabular}{cccccccccc}
				$J^P$ & $l$ & $L$ & $L_t$ & $n_{\max}$ & $\nu_1$ & $q$ & $N_{\max}$ & $\lambda_1$ & $Q$ \\ \hline
				$0^+$ &  0  &  0  &   0   &     20     &   20    & 0.6 &     20     &     30      & 0.6 \\
				$2^+$ &  2  &  0  &   2   &     20     &   20    & 0.6 &     20     &     30      & 0.6 \\
				$4^+$ &  4  &  0  &   4   &     20     &   20    & 0.6 &     20     &     30      & 0.6
			\end{tabular}
		\end{ruledtabular}
	\end{table}
 
    {
    To accelerate convergence, we employ three Jacobian coordinates for the $0^+$ state, namely
    \begin{equation}\label{eq:Phi2}
    \begin{aligned}
        \Phi_\alpha=&C_1\left([\phi_{nl}^{\mathrm{G}}(\boldsymbol{r}_1)\phi_{NL}^{\mathrm{G}}(\boldsymbol{R}_1)]_{L_tM_t}
        +[\phi_{nl}^{\mathrm{G}}(\boldsymbol{r}_2)\phi_{NL}^{\mathrm{G}}(\boldsymbol{R}_2)]_{L_tM_t}\right)\\
        &+C_2[\phi_{nl}^{\mathrm{G}}(\boldsymbol{r}_3)\phi_{NL}^{\mathrm{G}}(\boldsymbol{R}_3)]_{L_tM_t}.
    \end{aligned}
    \end{equation}
    Using the coordinate transformation 
        \begin{equation}
    		\begin{cases}
    			\boldsymbol{r}_a=\alpha_{ac}\boldsymbol{r}_c+\beta_{ac}\boldsymbol{R}_c,\\
    			\boldsymbol{R}_a=\gamma_{ac}\boldsymbol{r}_c+\delta_{ac}\boldsymbol{R}_c,
    		\end{cases}
    	\end{equation}
    the Gaussian basis can be rewritten as
    \begin{equation}\label{eq:coortrans}
    \begin{aligned}
        &\mathrm{e}^{-(\nu_{n_a}r_a^2+\lambda_{N_a}R_a^2)}=\mathrm{e}^{-(Ar_c^2+BR_c^2+2C\boldsymbol{r}_c\boldsymbol{\cdot}\boldsymbol{R}_c)}\\
        &=\mathrm{e}^{-(Ar_c^2+BR_c^2)}\sum_{l=0}^{\infty}4\pi\sqrt{2l+1}i_l(2Cr_cR_c)[Y_l(\hat{\boldsymbol{r}}_c)Y_l(\hat{\boldsymbol{R}_c})]_{00},
    \end{aligned}
    \end{equation}
    where $i_l(z)$ is the modified Bessel function. It shows that a basis with $(l,L)=(0,0)$ in the $c=1$ coordinates corresponds not only to a $(0,0)$ basis in the $c=2$ or $c=3$ coordinates (with different Gaussian ranges) but also contains higher‑angular‑momentum components such as $(1,1)$, $(2,2)$, etc. Hence, using three Jacobian coordinates allows us to incorporate contributions from these higher partial waves $(1,1)$, $(2,2)$, $\cdots$.
    
    The effect is illustrated for $\prescript{8}{\phi}{\mathrm{Be}}(0^+_1)$ using the $V_{\alpha\phi}^{\text{ave}}(r)$ potential with an $\alpha$-particle rms matter radius of 1.70 fm. Table~\ref{tab:Proportion} lists the energies obtained with different bases, together with the proportion of each basis. Note that bases with odd $l$ do not contribute to the system under discussion, as explained in detail in Ref.~\cite{Zhou:2025fpp}. One can see that when only a single Jacobian coordinate is used, the $(2,2)$ basis does give a non‑negligible contribution; however, convergence is significantly faster when three Jacobian coordinates are employed. It should be noted that the $\alpha$-$\alpha$ potential is known only for $l=0$, $2$, $4$; for other partial waves we use the $l=0$ potential as a substitute, which may slightly overestimate the energy levels of the system.

    \begin{table*}[htbp]
        \centering
        \caption{Energy levels of $\prescript{8}{\phi}{\mathrm{Be}}$ obtained using different basis sets with the $V_{\alpha\phi}^{\text{ave}}(r)$ potential, together with the weight of each basis.
    Due to the non‑orthogonality of the basis functions defined in different Jacobi coordinates, the last two entries in the second column give the projection of each basis onto the total wave function.
    The other Gaussian basis parameters are the same as those listed in Table~\ref{tab:Angular-momentum space}.
    The boldface energy levels correspond to the final values adopted in the article. NBS represents that there are no bound states. \label{tab:Proportion}}
        \begin{tabular}{ccc}
        \hline\hline
        $0^+_1$&$2^+_1$&$4^+_1$\\
        \hline
        \begin{tabular}[t]{crc}
            $(c,l,L)$&Weight&$E$ (MeV)\\
            \hline
            $(3,0,0)$&100.0&$-$12.0\\[10pt]
            $(3,0,0)$&100.0&\multirow{2}{*}{$-$12.0}\\
            $(3,1,1)$&0.00&\\[10pt]
            $(3,0,0)$&94.96&\\
            $(3,1,1)$&0.00&$-$13.9\\
            $(3,2,2)$&5.04&\\[10pt]
            $(1,0,0)$&\multirow{2}{*}{396.30}&\\
            $(2,0,0)$&&$\boldsymbol{-14.1}$\\
            $(3,0,0)$&$-$296.30&\\
        \end{tabular}&
        \begin{tabular}[t]{crc}
            $(c,l,L)$&Weight&$E$ (MeV)\\
            \hline
            $(3,2,0)$&100.0&$\boldsymbol{-9.92}$\\[10pt]
            $(3,0,2)$&100.0&NBS\\[10pt]
            $(3,2,0)$&99.04&\multirow{1.5}{*}{$-$10.3}\\
            $(3,0,2)$& 0.96&\\[10pt]
            $(1,2,0)$&\multirow{1.5}{*}{1.04}&\\
            $(2,2,0)$&&$-$9.99\\
            $(3,2,0)$&98.96&\\
        \end{tabular}&
        \begin{tabular}[t]{crc}
            $(c,l,L)$&Weight&$E$ (MeV)\\
            \hline
            $(3,4,0)$&100.0&$\boldsymbol{-6.12}$\\[10pt]
            $(3,2,2)$&100.0&NBS\\[10pt]
            $(3,0,4)$&100.0&NBS\\[10pt]
            $(3,4,0)$&99.01&\\
            $(3,2,2)$& 0.98&$-6.47$\\
            $(3,0,4)$& 0.01&\\
        \end{tabular}\\
        \hline\hline
        \end{tabular}
    \end{table*}

To constrain the orbital angular momentum between the two $\alpha$ particles, we employ only the third Jacobi coordinate for the $2^+$ and $4^+$ states.
For the $2^+$ state, besides the $(l,L)=(2,0)$ basis, the $(0,2)$ configuration is also available.
For the $4^+$ state, in addition to the $(4,0)$ basis, the $(2,2)$ and $(0,4)$ bases may be selected.
Based on previous studies~\cite{Lee:2019mlt, Wu:2019ivs}, which indicate that the $M$ particles predominantly occupy $S$-waves, we have adopted an $S$-wave angular momentum configuration for the $M$ particle.

Taking $\prescript{8}{\phi}{\mathrm{Be}}$ as an example, we evaluated the contributions of these bases using the $V_{\alpha\phi}^{\text{ave}}(r)$ potential, as summarized in Table~\ref{tab:Proportion}.
As expected, the contributions from these bases are very small. Moreover, when used alone, none of them yields a bound state. This is because orbital excitation along $\boldsymbol{R}$ introduces a centrifugal barrier that cancels the attraction between $\alpha$ and $\phi$.

Similar to the $S$-wave case, only $\alpha$-$\alpha$ potentials for $l=0,2,4$ are known; here we use the $l=0$ potential as a representative for other partial waves. For the reasons outlined above, we restrict our calculation to the third Jacobi coordinate and consider only excitation in the relative coordinate $l$—an approach that constitutes a reasonable approximation.
}
	
 The explicit form of the Gaussian basis functions in Eq.~\eqref{eq:Phi2} is
	\begin{equation}
		\begin{array}{ll}
			\phi_{nlm}^{\mathrm{G}}(\boldsymbol{r})=\phi_{nl}^{\mathrm{G}}(r)Y_{lm}(\hat{\boldsymbol{r}}),&\phi_{nl}^{\mathrm{G}}(r)=N_{nl}r^l\mathrm{e}^{-\nu_nr^2},\\
			\phi_{NLM}^{\mathrm{G}}(\boldsymbol{R})=\phi_{NL}^{\mathrm{G}}(R)Y_{LM}(\hat{\boldsymbol{R}}),&\phi_{NL}^{\mathrm{G}}(R)=N_{NL}R^L\mathrm{e}^{-\lambda_NR^2},
		\end{array}
	\end{equation}
	with the normalization constants
	\begin{equation}
		N_{nl}=\left(\frac{2^{l+2}(2\nu_n)^{l+\frac{3}{2}}}{\sqrt{\pi}(2l+1)!!}\right)^{\frac{1}{2}},\quad N_{NL}=\left(\frac{2^{L+2}(2\lambda_N)^{L+\frac{3}{2}}}{\sqrt{\pi}(2L+1)!!}\right)^{\frac{1}{2}},
	\end{equation}
	 and the Gaussian range parameters chosen in a geometric progression
	\begin{equation}
		\begin{array}{ll}
			\nu_n=\nu_1q^{n-1}&(n=1\sim n_{\max}),\\
			\lambda_N=\lambda_1Q^{N-1}&(N=1\sim N_{\max}).
		\end{array}
	\end{equation}
	Empirical experience shows that a geometric progression of Gaussian parameters provides an efficient and accurate basis set.
	
	\subsection{Complex scaling method}
	The CSM enables the calculation of resonant states by applying a complex rotation to the Jacobi coordinates:
	\begin{equation}
		r_c\to r_c\mathrm{e}^{\mathrm{i}\theta},\quad R_c\to R_c\mathrm{e}^{\mathrm{i}\theta}.
	\end{equation}
	 Under this transformation, the kinetic and potential terms of the Hamiltonian become
	\begin{equation}
		T\to T\mathrm{e}^{-2\mathrm{i}\theta},\quad V_{ij}(r_{ij})\to V_{ij}(r_{ij}\mathrm{e}^{\mathrm{i}\theta}).
	\end{equation}
	
	Solving the complex-scaled stationary Schr\"odinger equation for a given scaling angle $\theta$ yields a set of complex energy eigenvalues. Among these, resonant states are identified by their independent of $\theta$, bound states remains stable on the negative real axis, and  continuum states rotate downwards by an angle of $2\theta$ from the real axis. The complex eigenvalue of a resonance can be written as
	\begin{equation}
		E=E_r-\mathrm{i}\Gamma_r/2,
	\end{equation}
	where $E_r$ and $\Gamma_r$ are the resonance energy and width, respectively. As with the ordinary stationary Schr\"odinger equation, the complex-scaled euqation is solved using the GEM.
	
	To investigate the spatial structure of the system, we calculate the root-mean-square (rms) radius for each state. For bound states, it is defined as
	\begin{equation}
		R=\sqrt{\langle\Psi|r_{ij}^2|\Psi\rangle}.
	\end{equation}
	For resonant states within the CSM framework, the rms radius is obtained by
	\begin{equation}
R=\operatorname{Re}\left[\sqrt{\frac{(\Psi(\theta)|r_{ij}^2\mathrm{e}^{2\mathrm{i}\theta}| \Psi(\theta))}{(\Psi(\theta)|\Psi(\theta))}}\right],
	\end{equation}
	where $\Psi(\theta)$ is the complex wave function obtained from the complex-scaled stationary Schr\"odinger equation with angle $\theta$. The round inner product is defined as~\cite{Romo:1968tcz}
	\begin{equation}
        (\psi|\phi)=\int\psi(\boldsymbol{r})\phi(\boldsymbol{r})\mathrm{d}\boldsymbol{r},
	\end{equation}
	without complex conjugation of the bra vector. For resonant states, the rms radius obtained from this product is generally complex, whereas for bound states it remains real. As the rotation angle $\theta$ increases, the rms radius converges to a stable value, in agreement with the discussion in Ref.~\cite{Homma:1997wtc}.
	
	\section{numerical results}\label{sec3}
 
To obtain more stable results, bound states were calculated by solving the Schr\"odinger equation using the GEM, while resonant states were obtained by solving the complex-scaled Schr\"odinger equation with the same method.

\begin{figure*}[htbp]
        \begin{tikzpicture}
			\begin{axis}[
				width=18.6cm, height=11cm,
				axis line style={thick},
				ylabel=$E$ (MeV),
                ylabel style={at={(axis cs:0,16)},rotate=-90},
				ymin=-19.99, ymax=15,
				extra y ticks={-16},       
				extra y tick labels={$-$20},
				ytick distance=5, minor y tick num=4,
				ytick style={thick, black, line cap=round},
				yticklabel style={/pgf/number format/.cd, fixed, precision=0, fixed zerofill},
				major tick length=0.13cm,
				minor tick length=0.07cm,
				xmin=-0.5, xmax=9.5,
				xtick={0,1,2,3,4,5,6,7,8,8.8},
				xticklabels={$\prescript{8}{}{\text{Be}}$,$\prescript{9}{\Lambda}{\text{Be}}$~\cite{Lee:2019mlt,Wu:2019ivs},$\prescript{8}{\phi}{\text{Be}}(V_{\alpha\phi}^{3/2})$,$\prescript{8}{\phi}{\text{Be}}(V_{\alpha\phi}^{\text{ave}})$,$\prescript{8}{\phi}{\text{Be}}(V_{\alpha\phi}^{1/2})$,$\prescript{8}{J\!/\!\psi}{\text{Be}}(V_{\alpha J\!/\!\psi}^{3/2})$,$\prescript{8}{J\!/\!\psi}{\text{Be}}(V_{\alpha J\!/\!\psi}^{\text{ave}})$,$\prescript{8}{J\!/\!\psi}{\text{Be}}(V_{\alpha J\!/\!\psi}^{1/2})$,$\prescript{8}{\eta_c}{\text{Be}}$,continuum},
				xtick style={draw=none},
				nodes near coords,
				nodes near coords align={above=-1.5pt},
				coordinate style/.condition={y==-0.156||y==-5.34||y==-0.826||y==1.94||y==-1.05||y==2.11||y==-1.54||y==-0.219||y==2.59||y==11.0||y==-22.1+4}{below=-1.5pt},
				coordinate style/.condition={y==-21.1+4}{left=14},
				]
				\addplot[mark=-, mark size=0.5cm, semithick, line cap=round,
				draw=none, point meta=explicit symbolic,
				]coordinates
				{
					(0,0.09)[0.09 (1E-4)]
					(0,2.90)[2.90 (1.3)]
					(0,11.6)[11.6 (3.1)]
					(1,-6.65)[$-$6.65]
					(1,-3.82)[$-$3.82]
					(1,3.2)[3.2 (0.78)]
					(2,-9.34)[$-$9.34]
					(2,-5.34)[$-$5.34]
					(2,-0.156)[$-$0.156]
                    (3,-14.1)[$-$14.1]
					(3,-9.92)[$-$9.92]
					(3,-6.12)[$-$6.12]
					(4,-22.1+4)[$-$22.1]
					(4,-21.1+4)[$-$21.1]
					(4,-20.6+4)[$-$20.6]
					(5,-0.826)[$-$0.826]
					(5,2.11)[2.11 (0.42)]
					(5,10.1)[10.1 (1.8)]
                    (6,-1.05)[$-$1.05]
					(6,1.94)[1.94 (0.33)]
					(6,9.86)[9.86 (1.7)]
					(7,-1.54)[$-$1.54]
					(7,1.57)[1.57 (0.19)]
					(7,9.24)[9.24 (1.3)]
					(8,-0.219)[$-$0.219]
					(8,2.59)[2.59 (0.80)]
					(8,11.0)[11.0 (2.4)]
				};
				\node at(axis cs:0.55,0.8){$0^+_1$};
				\node at(axis cs:0.5,3.4){$2^+_1$};
				\node at(axis cs:0.5,12.2){$4^+_1$};
				\node at(axis cs:1.42,-6.2){$0^+_1$};
				\node at(axis cs:1.42,-3.4){$2^+_1$};
				\node at(axis cs:1.50,3.7){$4^+_1$};
				\node at(axis cs:2.4,-9){$0^+_1$};
				\node at(axis cs:2.4,-6.2){$2^+_1$};
				\node at(axis cs:2.4,-1){$4^+_1$};
				\node at(axis cs:3.4,-13.6){$0^+_1$};
				\node at(axis cs:3.4,-9.4){$2^+_1$};
				\node at(axis cs:3.4,-5.7){$4^+_1$};
				\node at(axis cs:4.4,-23+4){$0^+_1$};
				\node at(axis cs:4.4,-21.5+4){$2^+_1$};
				\node at(axis cs:4.4,-20+4){$4^+_1$};
				\node at(axis cs:4.53,-1.5){$0^+_1$};
				\node at(axis cs:4.48,1.4){$2^+_1$};
				\node at(axis cs:4.55,10.7){$4^+_1$};
				\node at(axis cs:5.6,-1.8){$0^+_1$};
				\node at(axis cs:5.52,1.2){$2^+_1$};
				\node at(axis cs:5.55,10.5){$4^+_1$};
				\node at(axis cs:6.6,-2.2){$0^+_1$};
				\node at(axis cs:6.5,2.2){$2^+_1$};
				\node at(axis cs:6.52,9.9){$4^+_1$};
				\node at(axis cs:7.6,-1){$0^+_1$};
				\node at(axis cs:7.5,1.65){$2^+_1$};
				\node at(axis cs:7.55,10.1){$4^+_1$};
				\draw[dashed] (axis cs:1.7,-4.59)--(axis cs:8.4,-4.59)node[right]{\footnotesize$\prescript{4}{\phi}{\text{He}}+\alpha$};
				\draw[dashed] (axis cs:-0.5,0)--(axis cs:8.4,0)node[right]{\footnotesize$\alpha+\alpha+M$};
				\draw[dashed] (axis cs:0.7,2.9)--(axis cs:8.4,2.9)node[right]{\footnotesize$\prescript{8}{}{\text{Be}}(2^+)+M$};
				\draw[dashed] (axis cs:0.7,11.6)--(axis cs:8.4,11.6)node[right]{\footnotesize$\prescript{8}{}{\text{Be}}(4^+)+M$};
			\end{axis}
		\end{tikzpicture}
		\caption{The energy spectra of $\prescript{8}{}{\text{Be}}$ from phenomenological $\alpha$-$\alpha$ potential and $\prescript{9}{\Lambda}{\text{Be}}$ from Ref.~\cite{Lee:2019mlt,Wu:2019ivs}, as well as the energy spectra of $\prescript{8}{\phi}{\text{Be}}$, $\prescript{8}{J/\psi}{\text{Be}}$ and $\prescript{8}{\eta_c}{\text{Be}}$ with an $\alpha$-particle rms matter radius of 1.70 fm. The values in parenthesis are decay widths, while the dashed lines represent the decay thresholds. Note that the spacing between $y$-axis $-$20 and $-$15 has been compressed for aesthetic purposes. {The content in parentheses after $\prescript{8}{M}{\text{Be}}$ refers to the $\alpha$-$M$ interaction used.} \label{fig:energy spectra}}
	\end{figure*}
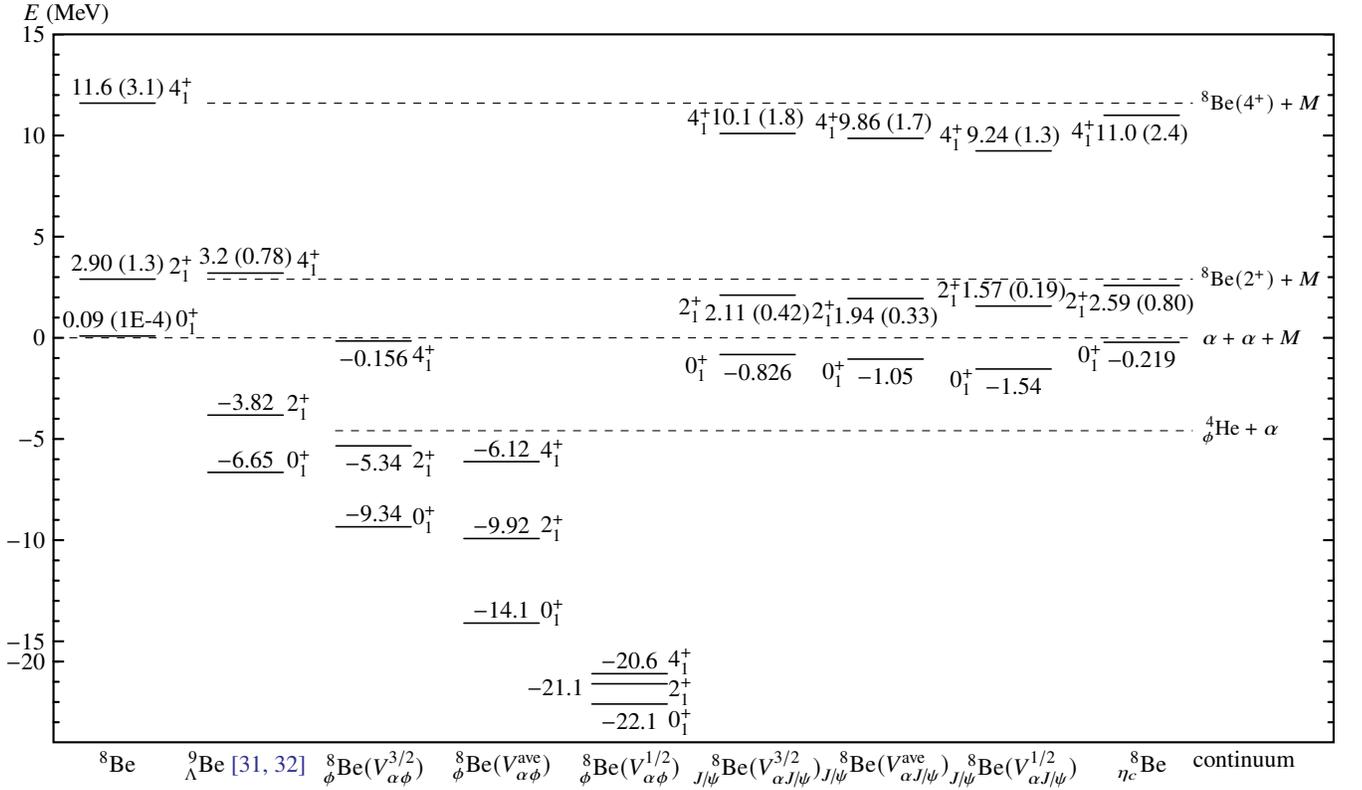

The energy spectra of $\prescript{8}{}{\text{Be}}$, $\prescript{9}{\Lambda}{\text{Be}}$, $\prescript{8}{\phi}{\text{Be}}$, $\prescript{8}{J/\psi}{\text{Be}}$, and $\prescript{8}{\eta_c}{\text{Be}}$ are shown in Fig.~\ref{fig:energy spectra}. The energy spectrum of $\prescript{8}{}{\text{Be}}$ was computed using a phenomenological $\alpha$-$\alpha$ potential with the CSM and GEM, and agrees well with the experimental values of 0.0918 MeV, 3.03 MeV, and 11.35 MeV for the $0^+_1$, $2^+_1$, and $4^+_1$ states, respectively~\cite{Tilley:2004zz}. The energy spectrum of $\prescript{9}{\Lambda}{\text{Be}}$ is taken from Refs.~\cite{Lee:2019mlt,Wu:2019ivs}, where it was derived using the OCM and CSM. The energy spectra of $\prescript{8}{\phi}{\text{Be}}$, $\prescript{8}{J/\psi}{\text{Be}}$, and $\prescript{8}{\eta_c}{\text{Be}}$ were calculated via GEM and CSM using the HAL QCD potential and a phenomenological $\alpha$-$\alpha$ potential, with the $\alpha$-particle root-mean-square matter radius $R_\alpha$ taken as 1.84 fm, 1.70 fm, and 1.56 fm, respectively. Fig.~\ref{fig:energy spectra} displays the case for $R_\alpha = 1.70$ fm.

Table~\ref{tab:energy spectra} presents the energy spectra of the $\alpha+M$ ($\prescript{4}{M}{\text{He}}$), $\alpha+\alpha$ ($\prescript{8}{}{\text{Be}}$), and $\alpha+\alpha+M$ ($\prescript{8}{M}{\text{Be}}$) systems, where $M$ denotes $\phi$, $J/\psi$, or $\eta_c$. To investigate the dynamic effects of the $M$ particle, we also calculated the rms radii $R_{\alpha M}$ and $R_{\alpha\alpha}$ for each system. The three values in each cell correspond to the $\alpha$-particle rms radius $R_\alpha$ taken as 1.84, 1.70 and 1.56~fm, respectively. From Table~\ref{tab:energy spectra}, the following general trends can be observed:
(1) Both the energy level $E$ and the radii $R_{\alpha M}$ and $R_{\alpha\alpha}$ decrease as $R_\alpha$ decreases.
(2) The energy level $E$ increases while the radii $R_{\alpha M}$ and $R_{\alpha\alpha}$ decrease with increasing orbital angular momentum.

 	\begin{table*}[htbp]
		\caption{The energy spectra of the $\alpha+M\ (\prescript{4}{M}{\text{He}})$, $\alpha+\alpha\ (\prescript{8}{}{\text{Be}})$, and $\alpha+\alpha+M\ (\prescript{8}{M}{\text{Be}})$ systems ($M$ stands for $\phi, J/\psi$, or $\eta_c$). The value in parentheses is the width of the resonant state. $R_{\alpha M}/R_{\alpha\alpha}$ is the rms radius of the corresponding state between $\alpha$ and $M/\alpha$. The three values in each cell correspond to the $\alpha$-particle rms radius $R_\alpha$ taken as 1.84, 1.70 and 1.56 fm, respectively. In addition, $R_\alpha$ used in Ref.~\cite{Etminan:2025yoq} is 1.47 fm. NBS represents that there are no bound states. \label{tab:energy spectra}}
		\begin{ruledtabular}
		\begin{tabular}{ccccccc}
			                       &         &                                       \multicolumn{3}{c}{$\alpha+M\ (\prescript{4}{M}{\text{He}})$}                                        &  \multicolumn{2}{c}{$\alpha+\alpha\ (\prescript{8}{}{\text{Be}})$}  \\ \cline{3-5}\cline{6-7}
			     Interaction       &         &               $E$ (MeV)               & Refs.~\cite{Filikhin:2024xkb,Etminan:2025yoq} & $R_{\alpha M}$ (fm) &  $E$ (MeV)  & $R_{\alpha\alpha}$ (fm) \\ \hline
			 $V_{\alpha\phi}^{3/2}(r)$  &         &          $-$3.40, $-$4.59, $-$6.18          &            $-$2.967, $-$4.780, $-$5.976             &      2.86, 2.55, 2.27      & 0.09 (1E-4) &          5.96           \\
    $V_{\alpha\phi}^{\text{ave}}(r)$  &         &          $-$6.14, $-$8.10, $-$10.7          &                    &      2.36, 2.12, 1.90      & 2.90 (1.3) &          3.60           \\
			 $V_{\alpha\phi}^{1/2}(r)$  &         &          $-$13.1, $-$17.0, $-$22.1          &                                               &      1.87, 1.69, 1.52      & 11.6 (3.1)  &          2.91           \\
			$V_{\alpha J/\psi}^{3/2}(r)$ &         &       $-$0.00654, $-$0.0424, $-$0.127       &                      NBS                      &      30.5, 12.5, 7.52      &   &                     \\
        $V_{\alpha J/\psi}^{\text{ave}}(r)$ &         &        $-$0.0375, $-$0.109, $-$0.244         &                     NBS                      &      13.3, 8.11, 5.65      &             &                         \\
			$V_{\alpha J/\psi}^{1/2}(r)$ &         &        $-$0.168, $-$0.322, $-$0.574         &                     $-$0.1                      &      6.76, 5.08, 3.97      &             &                         \\
			$V_{\alpha\eta_c}(r)$ &         &             NBS, NBS, NBS             &                      NBS                      &       NBS, NBS, NBS        &             &                         \\ \hline
			                       &                                                             \multicolumn{6}{c}{$\alpha+\alpha+M\ (\prescript{8}{M}{\text{Be}})$}                                                             \\ \cline{2-7}
			     Interaction       &  State  &               $E$ (MeV)               & Refs.~\cite{Filikhin:2024xkb,Etminan:2025yoq} & $R_{\alpha M}$ (fm) &             & $R_{\alpha\alpha}$ (fm) \\ \hline
			 $V_{\alpha\phi}^{3/2}(r)$  & $0^+_1$ &          $-$7.81, $-$9.34, $-$11.2          &             $-$7.03, $-$9.79, $-$10.86              &      3.09, 2.96, 2.85      &             &    3.63, 3.57, 3.51     \\
			                       & $2^+_1$ &          $-$4.06, $-$5.34, $-$6.87          &                                               &      2.93, 2.77, 2.62      &             &    3.38, 3.28, 3.17     \\
			                       & $4^+_1$ &      1.75 (0.012), $-$0.156, $-$2.51      &                                               &      3.13, 2.43, 2.26      &             &    3.39, 2.77, 2.63     \\[10pt]
			 $V_{\alpha\phi}^{\text{ave}}(r)$  & $0^+_1$ &          $-$11.8, $-$14.1, $-$16.9          &                                               &      2.84, 2.74, 2.65      &             &    3.49, 3.42, 3.36     \\
			                       & $2^+_1$ &          $-$7.96, $-$9.92, $-$12.2          &                                               &      2.63, 2.50, 2.36      &             &    3.19, 3.09, 2.98     \\
			                       & $4^+_1$ &          $-$3.19, $-$6.12, $-$9.72          &                                               &      2.32, 2.17, 2.03      &             &    2.68, 2.56, 2.45     \\[10pt]
			 $V_{\alpha\phi}^{1/2}(r)$  & $0^+_1$ &          $-$19.0, $-$22.1, $-$25.7          &                                               &      2.40, 2.28, 2.17      &             &    3.12, 3.03, 2.92     \\
			                       & $2^+_1$ &          $-$17.4, $-$21.1, $-$25.5          &                                               &      2.28, 2.16, 2.03      &             &    2.94, 2.83, 2.71     \\
			                       & $4^+_1$ &          $-$15.1, $-$20.6, $-$27.3          &                                               &      2.00, 1.87, 1.74      &             &    2.44, 2.33, 2.22     \\[10pt]
			$V_{\alpha J/\psi}^{3/2}(r)$ & $0^+_1$ &         $-$0.635, $-$0.826, $-$1.07         &                     $-$1.13                     &      4.48, 4.20, 3.95      &             &    4.50, 4.42, 4.33     \\
			                       & $2^+_1$ & 2.22 (0.51), 2.11 (0.42), 1.98 (0.33) &                                               &      4.28, 4.12, 3.98      &             &    3.92, 3.98, 4.03     \\
			                       & $4^+_1$ &  10.4 (2.1), 10.2 (1.8), 9.86 (1.6)   &                                               &      3.28, 3.11, 2.97      &             &    2.95, 2.96, 2.97     \\[10pt]
            $V_{\alpha J/\psi}^{\text{ave}}(r)$ & $0^+_1$ &          $-$0.82, $-$1.05, $-$1.34          &                     $-$1.31                     &      4.24, 3.99, 3.77      &             &    4.42, 4.34, 4.26     \\
			                       & $2^+_1$ & 2.07 (0.42), 1.94 (0.33), 1.79 (0.25) &                                               &      4.13, 3.99, 3.87      &             &    4.01, 4.07, 4.12     \\
			                       & $4^+_1$ &  10.1 (1.9), 9.86 (1.7), 9.53 (1.4)   &                                               &      3.15, 3.00, 2.87      &             &    2.96, 2.97, 2.98     \\[10pt]
			$V_{\alpha J/\psi}^{1/2}(r)$ & $0^+_1$ &          $-$1.23, $-$1.54, $-$1.94          &                     $-$1.71                     &      3.88, 3.68, 3.51      &             &    4.29, 4.22, 4.15     \\
			                       & $2^+_1$ & 1.75 (0.26), 1.57 (0.19), 1.35 (0.12) &                                               &      3.92, 3.82, 3.69      &             &    4.17, 4.23, 4.27     \\
			                       & $4^+_1$ &  9.60 (1.6), 9.24 (1.3), 8.81 (1.1)   &                                               &      2.96, 2.84, 2.73      &             &    2.98, 3.00, 3.02     \\[10pt]
			$V_{\alpha\eta_c}(r)$ & $0^+_1$ &        $-$0.135, $-$0.219, $-$0.334         &                     $-$0.56                     &      6.35, 5.71, 5.16      &             &    4.89, 4.78, 4.67     \\
			                       & $2^+_1$ & 2.63 (0.87), 2.59 (0.80), 2.53 (0.72) &                                               &      5.41, 5.09, 4.81      &             &    3.52, 3.59, 3.66     \\
			                       & $4^+_1$ &  11.1 (2.6), 11.0 (2.4), 10.9 (2.2)   &                                               &      4.13, 3.83, 3.57      &             &    2.93, 2.93, 2.93
		\end{tabular}
		\end{ruledtabular}
	\end{table*}
    
	\subsection{The $\prescript{8}{\phi}{\text{Be}}$ system}
    Using the same Woods-Saxon potential and parameters as Ref.~\cite{Filikhin:2024xkb}, we are able to reproduce the results of Ref.~\cite{Filikhin:2024xkb} using GEM. The results shown in Table~\ref{tab:energy spectra} are obtained using the analytical expressions in~\eqref{eq:analytical alpha-qq}, and slight discrepancies with Ref.~\cite{Filikhin:2024xkb} can be observed for $\prescript{4}{\phi}{\text{He}}$ and $\prescript{8}{\phi}{\text{Be}}$. These differences arise from subtle variations in the fitting procedures for the $N$-$\phi$ potential~\cite{Lyu:2022imf}. Furthermore, while Ref.~\cite{Filikhin:2024xkb} considered only the $V_{\alpha\phi}^{3/2}$ potential for the bound states of $\prescript{8}{\phi}{\text{Be}}$, our analysis also includes the $V_{\alpha\phi}^{1/2}$ and $V_{\alpha\phi}^{\text{ave}}$ interactions.

As shown in Fig.~\ref{fig:energy spectra}, owing to the glue-like role of $\Lambda$ hyperons, the $0^+_1$ and $2^+_1$ resonant states of $^8$Be are transformed into bound states. The glue-like effect of the $\phi$ meson is even stronger, binding the $0^+_1$, $2^+_1$, and $4^+_1$ resonant states of $^8$Be into bound states. The energy level of $\prescript{8}{\phi}{\text{Be}}$ obtained from the $V_{\alpha\phi}^{1/2}$ interaction is notably low, reflecting the strong attraction between $N$ and $\phi$ in this channel. In addition, the $0^+_1$, $2^+_1$, and $4^+_1$ states exhibit closely spaced energy levels, indicating that the $\alpha$-$\phi$ interaction dominates the $\prescript{8}{\phi}{\text{Be}}$ system, while the contribution of the $\alpha$-$\alpha$ interaction is minimal. Moreover, as shown in Table~\ref{tab:energy spectra}, the introduction of the $\phi$ meson reduces $R_{\alpha\alpha}$ for the $0^+_1$, $2^+_1$, and $4^+_1$ states compared to $^8$Be, underscoring the glue-like role of the $\phi$ meson in compressing the nucleus. The $R_{\alpha\phi}$ in $\prescript{8}{\phi}{\text{Be}}$ is generally larger than that in $\prescript{4}{\phi}{\text{He}}$, a consequence of the repulsion between the two $\alpha$ particles. 

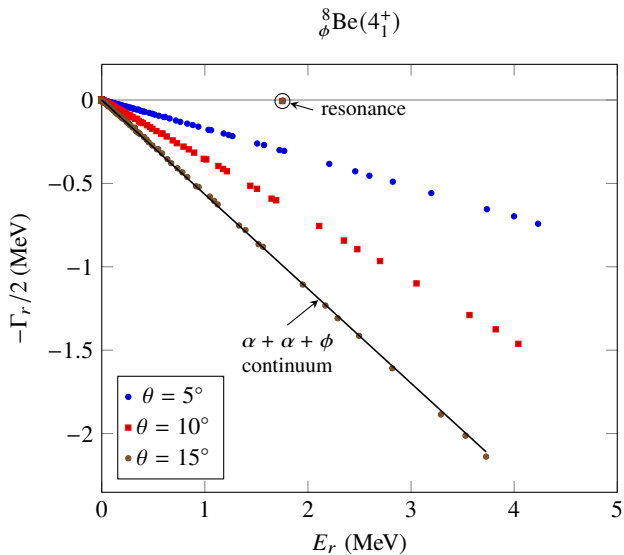
\begin{figure}[htbp]
		\begin{tikzpicture}
			\begin{axis}[
                title=$\prescript{8}{\phi}{\text{Be}}(4_1^+)$,
				xlabel=$E_r$ (MeV),
				xmin=0, xmax=5, 
				ylabel=$-\Gamma_r/2$ (MeV),
				legend entries={$\theta=5^\circ$,$\theta=10^\circ$,$\theta=15^\circ$},
				legend pos=south west,
				mark size=1pt,
				set layers,
				]
				\addplot+[only marks]coordinates{
					(1.43955e-08,-2.44273e-09)
					(4.20501e-08,-7.11921e-09)
					(7.20095e-08,-1.23268e-08)
					(1.09752e-07,-1.85233e-08)
					(1.78174e-07,-3.03796e-08)
					(2.67636e-07,-4.56039e-08)
					(2.78049e-07,-4.72092e-08)
					(4.23896e-07,-7.19352e-08)
					(6.02762e-07,-1.02994e-07)
					(6.67283e-07,-1.12084e-07)
					(8.79465e-07,-1.51566e-07)
					(9.98193e-07,-1.68605e-07)
					(1.33206e-06,-2.26603e-07)
					(1.54617e-06,-2.58778e-07)
					(1.86252e-06,-3.19415e-07)
					(2.3188e-06,-3.8966e-07)
					(2.5931e-06,-4.4749e-07)
					(2.91283e-06,-4.93021e-07)
					(3.40935e-06,-5.69056e-07)
					(3.93815e-06,-6.71985e-07)
					(5.20714e-06,-8.77169e-07)
					(5.3583e-06,-9.13818e-07)
					(6.26765e-06,-1.0531e-06)
					(7.09297e-06,-1.20065e-06)
					(7.16008e-06,-1.21749e-06)
					(8.26246e-06,-1.40336e-06)
					(1.07588e-05,-1.83053e-06)
					(1.14016e-05,-1.90938e-06)
					(1.29134e-05,-2.13999e-06)
					(1.43173e-05,-2.43768e-06)
					(1.44242e-05,-2.4308e-06)
					(1.70294e-05,-2.88092e-06)
					(1.85564e-05,-3.20454e-06)
					(2.16974e-05,-3.66755e-06)
					(2.31036e-05,-3.83747e-06)
					(2.3557e-05,-3.88943e-06)
					(2.84494e-05,-4.85565e-06)
					(2.95546e-05,-4.92784e-06)
					(3.4278e-05,-5.77113e-06)
					(3.67588e-05,-6.3082e-06)
					(3.75316e-05,-6.1209e-06)
					(4.24788e-05,-7.08189e-06)
					(4.49382e-05,-7.54681e-06)
					(4.62993e-05,-7.98588e-06)
					(5.49213e-05,-9.18164e-06)
					(5.67568e-05,-9.4267e-06)
					(6.14585e-05,-1.01645e-05)
					(6.74505e-05,-1.13024e-05)
					(7.10843e-05,-1.21264e-05)
					(7.60897e-05,-1.25909e-05)
					(8.76982e-05,-1.46657e-05)
					(9.01959e-05,-1.54776e-05)
					(0.000103085,-1.71741e-05)
					(0.000109293,-1.8462e-05)
					(0.000111097,-1.89104e-05)
					(0.000116322,-1.93405e-05)
					(0.000121342,-2.02217e-05)
					(0.000133234,-2.23045e-05)
					(0.000137148,-2.30061e-05)
					(0.000164175,-2.71019e-05)
					(0.000169931,-2.8943e-05)
					(0.000181397,-3.01515e-05)
					(0.00019792,-3.3086e-05)
					(0.000203718,-3.4242e-05)
					(0.000213325,-3.65519e-05)
					(0.000241169,-3.99843e-05)
					(0.000251126,-4.2148e-05)
					(0.000256337,-4.25502e-05)
					(0.000261204,-4.453e-05)
					(0.000311496,-5.27176e-05)
					(0.000324536,-5.399e-05)
					(0.000338616,-5.6126e-05)
					(0.000370291,-6.18112e-05)
					(0.00038955,-6.63307e-05)
					(0.000405821,-6.75609e-05)
					(0.00044925,-7.43844e-05)
					(0.000469334,-7.82862e-05)
					(0.000486464,-8.31663e-05)
					(0.000521312,-8.50479e-05)
					(0.000549893,-9.18437e-05)
					(0.000566311,-9.60447e-05)
					(0.000601575,-0.000101854)
					(0.000653714,-0.00010803)
					(0.000687131,-0.000116217)
					(0.000696112,-0.000117575)
					(0.000795408,-0.00013079)
					(0.000822784,-0.000136124)
					(0.000861822,-0.00014613)
					(0.000890779,-0.000148156)
					(0.000981628,-0.000164448)
					(0.00107401,-0.000183023)
					(0.00108806,-0.000180005)
					(0.00119138,-0.000200585)
					(0.00119908,-0.000202738)
					(0.00127877,-0.000218287)
					(0.00136446,-0.000225495)
					(0.00140726,-0.0002326)
					(0.00149666,-0.000252078)
					(0.00154752,-0.000257122)
					(0.00167036,-0.000276514)
					(0.00177445,-0.000297685)
					(0.00183864,-0.000311155)
					(0.00204173,-0.000338385)
					(0.00209024,-0.000348377)
					(0.00214134,-0.00035814)
					(0.00222096,-0.000367684)
					(0.00236272,-0.000398018)
					(0.0024449,-0.000410381)
					(0.00274845,-0.000465535)
					(0.00284624,-0.000469495)
					(0.00297792,-0.000498048)
					(0.00303635,-0.000499412)
					(0.00336137,-0.000559207)
					(0.00347616,-0.00057562)
					(0.00364351,-0.000610457)
					(0.00387869,-0.000652642)
					(0.00401152,-0.000665087)
					(0.00413188,-0.000715479)
					(0.00443359,-0.000713026)
					(0.00447047,-0.000743093)
					(0.00480375,-0.000794149)
					(0.00493318,-0.0008232)
					(0.00511928,-0.000855322)
					(0.00582422,-0.000960854)
					(0.005908,-0.00100293)
					(0.00620279,-0.000992442)
					(0.00623642,-0.00102303)
					(0.00664133,-0.00110353)
					(0.00693967,-0.00115127)
					(0.00768038,-0.00128372)
					(0.00801389,-0.00128719)
					(0.00834763,-0.00129952)
					(0.00878322,-0.00146333)
					(0.00938869,-0.00156034)
					(0.00980163,-0.00164472)
					(0.010094,-0.00172637)
					(0.0102433,-0.00167378)
					(0.010787,-0.00167278)
					(0.0111476,-0.00189278)
					(0.0115951,-0.00196598)
					(0.0120178,-0.00200457)
					(0.012576,-0.00207678)
					(0.0134293,-0.00222458)
					(0.0139063,-0.00219721)
					(0.0148019,-0.00243911)
					(0.0149996,-0.00246891)
					(0.0161921,-0.00269062)
					(0.0175409,-0.00276945)
					(0.0183124,-0.00297987)
					(0.0193392,-0.00314023)
					(0.0196147,-0.0032491)
					(0.0220263,-0.00358689)
					(0.0221338,-0.00370005)
					(0.0226927,-0.00382423)
					(0.0236801,-0.00410686)
					(0.023868,-0.00376256)
					(0.0251165,-0.00402859)
					(0.0271851,-0.00452214)
					(0.0279629,-0.00476744)
					(0.0294342,-0.00485449)
					(0.0307488,-0.00509568)
					(0.032134,-0.00514122)
					(0.033882,-0.00556751)
					(0.0343119,-0.00558295)
					(0.0370968,-0.0061355)
					(0.0406971,-0.00671646)
					(0.0418823,-0.00701272)
					(0.0424584,-0.0068099)
					(0.0441639,-0.00719576)
					(0.0448611,-0.00721204)
					(0.049568,-0.00825608)
					(0.0523027,-0.00899911)
					(0.0572018,-0.00961027)
					(0.0576273,-0.00927941)
					(0.0591419,-0.00951384)
					(0.0609137,-0.00994083)
					(0.0633045,-0.0107201)
					(0.0707181,-0.011755)
					(0.0711914,-0.0119971)
					(0.0747232,-0.0123709)
					(0.0762647,-0.0125806)
					(0.0833944,-0.0137462)
					(0.0863098,-0.0142614)
					(0.0869456,-0.0143872)
					(0.0958052,-0.0158563)
					(0.105971,-0.0175712)
					(0.107635,-0.0176819)
					(0.108996,-0.0183758)
					(0.113339,-0.0183277)
					(0.122604,-0.0203903)
					(0.128662,-0.0217668)
					(0.129622,-0.0217006)
					(0.142592,-0.0236061)
					(0.146909,-0.0247984)
					(0.157627,-0.0266743)
					(0.163612,-0.0277478)
					(0.169734,-0.027884)
					(0.171192,-0.0285334)
					(0.181072,-0.0300779)
					(0.187908,-0.0314368)
					(0.203473,-0.0338524)
					(0.213106,-0.0359505)
					(0.222373,-0.0376847)
					(0.230687,-0.0388264)
					(0.245766,-0.0403306)
					(0.271207,-0.0448333)
					(0.278852,-0.046715)
					(0.28692,-0.0475914)
					(0.30863,-0.0522372)
					(0.309587,-0.0518439)
					(0.339006,-0.0572285)
					(0.356632,-0.0573853)
					(0.376987,-0.0648882)
					(0.396188,-0.0671653)
					(0.399532,-0.0680752)
					(0.416298,-0.0704113)
					(0.458727,-0.0766064)
					(0.489448,-0.082155)
					(0.513812,-0.0877883)
					(0.554142,-0.0945253)
					(0.602016,-0.102256)
					(0.613627,-0.101699)
					(0.657728,-0.110887)
					(0.719268,-0.122895)
					(0.761806,-0.131382)
					(0.82896,-0.141773)
					(0.882474,-0.150174)
					(0.937225,-0.160052)
					(1.03504,-0.178928)
					(1.06012,-0.18072)
					(1.18239,-0.200625)
					(1.22963,-0.20964)
					(1.26767,-0.216631)
					(1.50765,-0.260875)
					(1.57389,-0.270029)
					(1.72425,-0.300129)
					(1.75427,-0.00383751)
					(1.76961,-0.305188)
					(2.20669,-0.38307)
					(2.45858,-0.427608)
					(2.59602,-0.45403)
					(2.8231,-0.489783)
					(3.19659,-0.558115)
					(3.73598,-0.654586)
					(3.9992,-0.69747)
					(4.2325,-0.742018)
					(5.86589,-1.02663)
				};
				\addplot+[only marks]coordinates{
					(1.37736e-08,-4.81369e-09)
					(4.02318e-08,-1.40302e-08)
					(6.8864e-08,-2.42886e-08)
					(1.05002e-07,-3.65067e-08)
					(1.70513e-07,-5.98595e-08)
					(2.56045e-07,-8.9871e-08)
					(2.65849e-07,-9.30307e-08)
					(4.06062e-07,-1.41727e-07)
					(5.76529e-07,-2.0296e-07)
					(6.38553e-07,-2.20929e-07)
					(8.40664e-07,-2.98616e-07)
					(9.56655e-07,-3.32224e-07)
					(1.2744e-06,-4.4659e-07)
					(1.48035e-06,-5.10134e-07)
					(1.78085e-06,-6.29378e-07)
					(2.22262e-06,-7.68046e-07)
					(2.47832e-06,-8.8169e-07)
					(2.78771e-06,-9.7174e-07)
					(3.26594e-06,-1.12182e-06)
					(3.76614e-06,-1.32433e-06)
					(4.98901e-06,-1.73003e-06)
					(5.12504e-06,-1.8e-06)
					(6.00219e-06,-2.0758e-06)
					(6.81434e-06,-2.36025e-06)
					(6.82575e-06,-2.40553e-06)
					(7.90348e-06,-2.76618e-06)
					(1.02947e-05,-3.60836e-06)
					(1.09226e-05,-3.76354e-06)
					(1.23786e-05,-4.21911e-06)
					(1.37137e-05,-4.82893e-06)
					(1.37978e-05,-4.76624e-06)
					(1.62951e-05,-5.67911e-06)
					(1.77344e-05,-6.31413e-06)
					(2.07719e-05,-7.2295e-06)
					(2.21479e-05,-7.56725e-06)
					(2.25822e-05,-7.66647e-06)
					(2.72128e-05,-9.56981e-06)
					(2.83262e-05,-9.71388e-06)
					(3.2817e-05,-1.13773e-05)
					(3.51486e-05,-1.24335e-05)
					(3.60146e-05,-1.20673e-05)
					(4.07085e-05,-1.39622e-05)
					(4.30364e-05,-1.48751e-05)
					(4.42509e-05,-1.57359e-05)
					(5.26587e-05,-1.81141e-05)
					(5.43529e-05,-1.85705e-05)
					(5.89269e-05,-2.00397e-05)
					(6.46019e-05,-2.22804e-05)
					(6.79972e-05,-2.38984e-05)
					(7.29499e-05,-2.48249e-05)
					(8.40167e-05,-2.89106e-05)
					(8.62417e-05,-3.04994e-05)
					(9.8789e-05,-3.38605e-05)
					(0.000104699,-3.63837e-05)
					(0.000106231,-3.72852e-05)
					(0.000111441,-3.81139e-05)
					(0.000116269,-3.98641e-05)
					(0.000127685,-4.39805e-05)
					(0.000131304,-4.53398e-05)
					(0.000157455,-5.34366e-05)
					(0.000162554,-5.70405e-05)
					(0.000173852,-5.94475e-05)
					(0.000189618,-6.52265e-05)
					(0.000195089,-6.74979e-05)
					(0.000203995,-7.20295e-05)
					(0.00023121,-7.88358e-05)
					(0.000240685,-8.3145e-05)
					(0.00024576,-8.37841e-05)
					(0.000249632,-8.7803e-05)
					(0.000298158,-0.000103907)
					(0.000311106,-0.000106463)
					(0.000324636,-0.000110668)
					(0.000354812,-0.000121863)
					(0.000372694,-0.000130727)
					(0.000388855,-0.000133178)
					(0.000430685,-0.000146653)
					(0.000449681,-0.000154334)
					(0.00046527,-0.0001639)
					(0.000500395,-0.000167737)
					(0.000526892,-0.000181068)
					(0.000542142,-0.000189352)
					(0.00057561,-0.000200715)
					(0.000626858,-0.000213014)
					(0.000657702,-0.000229056)
					(0.000666378,-0.000231749)
					(0.000762929,-0.000257893)
					(0.000788754,-0.00026838)
					(0.000824893,-0.000288078)
					(0.000853637,-0.000292054)
					(0.000940325,-0.00032421)
					(0.00102758,-0.000360867)
					(0.00104324,-0.000354798)
					(0.00114116,-0.000395476)
					(0.00114756,-0.00039958)
					(0.00122323,-0.000430275)
					(0.00130824,-0.000444595)
					(0.00135003,-0.000458818)
					(0.00143307,-0.00049692)
					(0.00148335,-0.000507014)
					(0.00160148,-0.000545226)
					(0.00169929,-0.000586756)
					(0.00175998,-0.000613381)
					(0.00195747,-0.000667275)
					(0.0020068,-0.000688046)
					(0.0020521,-0.000706641)
					(0.00212587,-0.000723839)
					(0.00226217,-0.000784515)
					(0.00234185,-0.000809274)
					(0.00263078,-0.00091787)
					(0.00272955,-0.000925732)
					(0.00285496,-0.000983411)
					(0.00291242,-0.000984149)
					(0.00322346,-0.00110379)
					(0.00333167,-0.00113493)
					(0.00349051,-0.00120356)
					(0.00371448,-0.00128663)
					(0.0038456,-0.00131123)
					(0.0039478,-0.00140977)
					(0.00426419,-0.00140801)
					(0.00428564,-0.00146607)
					(0.00460643,-0.00156552)
					(0.0047268,-0.00162327)
					(0.00490368,-0.00168606)
					(0.00559066,-0.00189597)
					(0.00565192,-0.00197808)
					(0.00598465,-0.00192524)
					(0.00596138,-0.0020495)
					(0.006367,-0.0021768)
					(0.0066511,-0.00226961)
					(0.00735979,-0.00253157)
					(0.00772478,-0.00255485)
					(0.00801591,-0.00254933)
					(0.00842389,-0.00288649)
					(0.00899776,-0.00307654)
					(0.00939142,-0.00324356)
					(0.00965094,-0.00340241)
					(0.00984096,-0.00330877)
					(0.0103929,-0.00328281)
					(0.0106531,-0.00374027)
					(0.0110991,-0.0038761)
					(0.011517,-0.00395328)
					(0.0120667,-0.00409606)
					(0.0128724,-0.00438983)
					(0.0133824,-0.00433118)
					(0.0141946,-0.00480709)
					(0.0143752,-0.00487058)
					(0.0155184,-0.00530465)
					(0.0168885,-0.00546499)
					(0.017579,-0.0058809)
					(0.0185711,-0.00618795)
					(0.0187871,-0.00640874)
					(0.0211803,-0.00705397)
					(0.0211892,-0.00732147)
					(0.0217307,-0.00754263)
					(0.0226191,-0.00810505)
					(0.0229564,-0.00740467)
					(0.0241358,-0.0079458)
					(0.0260502,-0.00891534)
					(0.0267511,-0.00939708)
					(0.0282438,-0.00957895)
					(0.029475,-0.0100452)
					(0.030889,-0.0101477)
					(0.0325549,-0.0110043)
					(0.0328706,-0.0109786)
					(0.0355583,-0.0120974)
					(0.039046,-0.0132569)
					(0.0401441,-0.0138615)
					(0.040756,-0.01339)
					(0.0424177,-0.014213)
					(0.043058,-0.0142018)
					(0.0475091,-0.0162774)
					(0.0499937,-0.0177331)
					(0.0548098,-0.0189778)
					(0.0553854,-0.0183123)
					(0.0568845,-0.0187454)
					(0.0583761,-0.0195999)
					(0.0605914,-0.021132)
					(0.0677898,-0.0231793)
					(0.0681615,-0.0236485)
					(0.0716077,-0.0243879)
					(0.073136,-0.024808)
					(0.0800102,-0.0271314)
					(0.0829605,-0.0281861)
					(0.0831799,-0.0283094)
					(0.0918524,-0.0312651)
					(0.101713,-0.0347079)
					(0.103123,-0.0348029)
					(0.104384,-0.0362673)
					(0.108853,-0.0361383)
					(0.117526,-0.040211)
					(0.12321,-0.0429661)
					(0.124118,-0.0427358)
					(0.136744,-0.0465567)
					(0.140647,-0.0488814)
					(0.150928,-0.0525995)
					(0.156738,-0.0548428)
					(0.163549,-0.0543475)
					(0.16336,-0.0568475)
					(0.173547,-0.0592993)
					(0.180024,-0.0620005)
					(0.19503,-0.0667577)
					(0.203981,-0.0708591)
					(0.212863,-0.0743331)
					(0.220917,-0.0765459)
					(0.23616,-0.0797167)
					(0.260104,-0.0884308)
					(0.267206,-0.0921494)
					(0.274997,-0.0938188)
					(0.2955,-0.103707)
					(0.296699,-0.101645)
					(0.32469,-0.112934)
					(0.34333,-0.113267)
					(0.360252,-0.127936)
					(0.379332,-0.132454)
					(0.382094,-0.134155)
					(0.398426,-0.138793)
					(0.439751,-0.151158)
					(0.468821,-0.161981)
					(0.491445,-0.173019)
					(0.530172,-0.186339)
					(0.576326,-0.20166)
					(0.588711,-0.200701)
					(0.629814,-0.218605)
					(0.688075,-0.242247)
					(0.728155,-0.258882)
					(0.793065,-0.279483)
					(0.844307,-0.295975)
					(0.896613,-0.315512)
					(0.989605,-0.352923)
					(1.01448,-0.356261)
					(1.1317,-0.395515)
					(1.17656,-0.413291)
					(1.21341,-0.427264)
					(1.44103,-0.514136)
					(1.5058,-0.532497)
					(1.64701,-0.591345)
					(1.75159,-0.00607791)
					(1.69162,-0.601416)
					(2.1095,-0.75544)
					(2.34879,-0.842574)
					(2.47895,-0.89456)
					(2.69905,-0.965583)
					(3.05295,-1.09967)
					(3.56674,-1.2895)
					(3.82088,-1.37498)
					(4.04095,-1.46194)
					(5.60176,-2.02306)
				};
				\addplot+[only marks]coordinates{
					(1.27602e-08,-7.04307e-09)
					(3.73038e-08,-2.05272e-08)
					(6.37125e-08,-3.55315e-08)
					(9.74724e-08,-5.34088e-08)
					(1.57812e-07,-8.75868e-08)
					(2.37321e-07,-1.31571e-07)
					(2.46393e-07,-1.36011e-07)
					(3.76004e-07,-2.07455e-07)
					(5.33655e-07,-2.96959e-07)
					(5.93091e-07,-3.23319e-07)
					(7.77228e-07,-4.3686e-07)
					(8.86467e-07,-4.86398e-07)
					(1.18028e-06,-6.53502e-07)
					(1.375e-06,-7.4675e-07)
					(1.64771e-06,-9.20927e-07)
					(2.06153e-06,-1.12451e-06)
					(2.29082e-06,-1.28971e-06)
					(2.58356e-06,-1.42213e-06)
					(3.0332e-06,-1.64244e-06)
					(3.48721e-06,-1.93799e-06)
					(4.62979e-06,-2.53237e-06)
					(4.74239e-06,-2.63387e-06)
					(5.5684e-06,-3.03859e-06)
					(6.33292e-06,-3.4376e-06)
					(6.30406e-06,-3.53703e-06)
					(7.32276e-06,-4.04817e-06)
					(9.53547e-06,-5.28002e-06)
					(1.01385e-05,-5.51e-06)
					(1.15055e-05,-6.17798e-06)
					(1.26922e-05,-7.08685e-06)
					(1.28005e-05,-6.95635e-06)
					(1.51052e-05,-8.31163e-06)
					(1.63916e-05,-9.23641e-06)
					(1.92573e-05,-1.05811e-05)
					(2.0596e-05,-1.10893e-05)
					(2.09806e-05,-1.12186e-05)
					(2.51938e-05,-1.40034e-05)
					(2.63061e-05,-1.42228e-05)
					(3.04376e-05,-1.66532e-05)
					(3.25148e-05,-1.81935e-05)
					(3.35412e-05,-1.76745e-05)
					(3.78137e-05,-2.04445e-05)
					(3.99284e-05,-2.17739e-05)
					(4.0904e-05,-2.30206e-05)
					(4.89597e-05,-2.65558e-05)
					(5.04334e-05,-2.71538e-05)
					(5.47672e-05,-2.93449e-05)
					(5.99573e-05,-3.26169e-05)
					(6.29536e-05,-3.49702e-05)
					(6.78189e-05,-3.6355e-05)
					(7.79999e-05,-4.23275e-05)
					(7.97805e-05,-4.46228e-05)
					(9.1768e-05,-4.95805e-05)
					(9.71952e-05,-5.32354e-05)
					(9.82776e-05,-5.46062e-05)
					(0.000103461,-5.57671e-05)
					(0.000107978,-5.83642e-05)
					(0.000118615,-6.44218e-05)
					(0.000121755,-6.63434e-05)
					(0.000146461,-7.82613e-05)
					(0.000150499,-8.34721e-05)
					(0.000161534,-8.70525e-05)
					(0.000176048,-9.5502e-05)
					(0.000180985,-9.88033e-05)
					(0.000188749,-0.000105389)
					(0.000214929,-0.000115456)
					(0.000223577,-0.000121875)
					(0.000228448,-0.00012245)
					(0.000230783,-0.000128563)
					(0.000276358,-0.000152085)
					(0.00028915,-0.000155945)
					(0.000301782,-0.000162092)
					(0.000329511,-0.000178436)
					(0.000345148,-0.000191307)
					(0.000361133,-0.000194963)
					(0.000400336,-0.00021475)
					(0.000417567,-0.000225961)
					(0.00043063,-0.000239832)
					(0.000466199,-0.000245778)
					(0.00048929,-0.000265126)
					(0.000502625,-0.000277259)
					(0.000533171,-0.000293667)
					(0.000582955,-0.000311995)
					(0.000609615,-0.000335238)
					(0.000617781,-0.000339212)
					(0.000709844,-0.000377727)
					(0.000733125,-0.00039301)
					(0.000764474,-0.000421757)
					(0.000792952,-0.000427553)
					(0.000872793,-0.000474721)
					(0.000951496,-0.000528274)
					(0.000970146,-0.000519527)
					(0.00105907,-0.000579099)
					(0.00106332,-0.000584811)
					(0.00113233,-0.000629776)
					(0.00121633,-0.000651087)
					(0.00125641,-0.000672451)
					(0.00132907,-0.00072746)
					(0.00137834,-0.000742602)
					(0.00148888,-0.00079854)
					(0.00157646,-0.000858851)
					(0.00163131,-0.000897935)
					(0.00181965,-0.000977386)
					(0.00187045,-0.00101076)
					(0.0019054,-0.00103574)
					(0.00197035,-0.00105704)
					(0.00209784,-0.00114827)
					(0.0021729,-0.0011853)
					(0.00243803,-0.00134389)
					(0.00253882,-0.00135583)
					(0.00265211,-0.00144329)
					(0.00271089,-0.00144017)
					(0.00299665,-0.00161906)
					(0.00309492,-0.00166162)
					(0.00324039,-0.00176244)
					(0.00344596,-0.0018838)
					(0.00357422,-0.00192001)
					(0.00364718,-0.00206224)
					(0.00398665,-0.0020675)
					(0.00398221,-0.00214855)
					(0.00428396,-0.00229195)
					(0.00438921,-0.00237766)
					(0.00455097,-0.00246814)
					(0.00520571,-0.00278016)
					(0.00523359,-0.00289711)
					(0.0056114,-0.00281517)
					(0.00552975,-0.00300907)
					(0.00591713,-0.00318974)
					(0.00617944,-0.00332296)
					(0.00683502,-0.00370818)
					(0.00723206,-0.00378945)
					(0.00749352,-0.003696)
					(0.00783638,-0.00423039)
					(0.00835815,-0.00450556)
					(0.00872025,-0.00475171)
					(0.00892608,-0.00497752)
					(0.00917377,-0.00486514)
					(0.00974775,-0.00478721)
					(0.00985572,-0.00548401)
					(0.0102876,-0.0056758)
					(0.0106971,-0.00579146)
					(0.0112348,-0.00600208)
					(0.0119555,-0.00643368)
					(0.012532,-0.00634434)
					(0.0131994,-0.00703423)
					(0.0133554,-0.00714018)
					(0.014417,-0.00776763)
					(0.0158204,-0.00801414)
					(0.0163781,-0.00862695)
					(0.0173138,-0.00905381)
					(0.0174372,-0.00939105)
					(0.0197739,-0.0103369)
					(0.0196635,-0.0107351)
					(0.0201524,-0.0110488)
					(0.0208816,-0.0118602)
					(0.0214754,-0.01084)
					(0.022534,-0.0116445)
					(0.0241945,-0.0130543)
					(0.0247707,-0.0137546)
					(0.0262944,-0.0140469)
					(0.0273943,-0.0147062)
					(0.0288484,-0.0148884)
					(0.0303333,-0.0162315)
					(0.0305683,-0.0159564)
					(0.0330428,-0.017717)
					(0.036333,-0.0194464)
					(0.0372627,-0.0203328)
					(0.038017,-0.019572)
					(0.0395357,-0.0208781)
					(0.0401364,-0.0207523)
					(0.0441426,-0.0238355)
					(0.0462213,-0.0259445)
					(0.0508569,-0.0278339)
					(0.051711,-0.0268576)
					(0.0532474,-0.0274393)
					(0.0542143,-0.0287159)
					(0.0561565,-0.0309356)
					(0.0630014,-0.0339519)
					(0.0632087,-0.0346175)
					(0.066506,-0.035702)
					(0.06802,-0.0363378)
					(0.0744522,-0.0397975)
					(0.0770632,-0.0417041)
					(0.0774443,-0.0410572)
					(0.085387,-0.0457896)
					(0.0946965,-0.0510462)
					(0.0957936,-0.0507556)
					(0.0967945,-0.0531736)
					(0.101544,-0.0529503)
					(0.10922,-0.0589073)
					(0.114202,-0.0629976)
					(0.115179,-0.0624885)
					(0.127176,-0.0682121)
					(0.130411,-0.0715535)
					(0.139957,-0.0770383)
					(0.145317,-0.0805225)
					(0.152866,-0.0793268)
					(0.15124,-0.0835224)
					(0.161237,-0.0868234)
					(0.167091,-0.0908177)
					(0.18121,-0.097788)
					(0.189065,-0.103711)
					(0.197233,-0.108896)
					(0.204935,-0.112076)
					(0.220332,-0.117208)
					(0.241897,-0.129551)
					(0.248124,-0.135049)
					(0.255514,-0.137354)
					(0.273579,-0.152067)
					(0.275877,-0.148901)
					(0.30114,-0.16554)
					(0.321594,-0.166224)
					(0.332891,-0.187334)
					(0.351724,-0.194009)
					(0.353562,-0.196291)
					(0.369195,-0.203161)
					(0.408595,-0.221594)
					(0.43507,-0.237191)
					(0.454878,-0.253199)
					(0.490948,-0.272777)
					(0.534243,-0.295393)
					(0.5479,-0.29434)
					(0.584145,-0.320037)
					(0.637042,-0.354569)
					(0.673174,-0.378731)
					(0.734355,-0.409156)
					(0.781914,-0.433138)
					(0.830153,-0.461859)
					(0.914959,-0.516915)
					(0.939836,-0.521558)
					(1.04876,-0.579037)
					(1.08976,-0.605112)
					(1.1246,-0.625977)
					(1.33211,-0.752363)
					(1.39436,-0.779944)
					(1.52081,-0.864982)
					(1.75242,-0.00635935)
					(1.56419,-0.879955)
					(1.95031,-1.1066)
					(2.16936,-1.2326)
					(2.28762,-1.30846)
					(2.49633,-1.41379)
					(2.81822,-1.60856)
					(3.29031,-1.88564)
					(3.52884,-2.01294)
					(3.72789,-2.13827)
					(5.16975,-2.9598)
				};
				
				\draw[gray,thin](axis cs:0,0)--(axis cs:5,0);
				
				\begin{pgfonlayer}{axis foreground}
					\draw[semithick](axis cs:0,0)--(axis cs:3.72789,-2.11);
					\draw[-stealth](axis cs:1.8,-1.35)node[below,align=center,font=\footnotesize]{$\alpha+\alpha+\phi$\\continuum}--(axis cs:2.1,-1.2);
					\node[circle,draw,inner sep=2](res)at(axis cs:1.75242,-0.00635935){};
					\draw[-stealth](axis cs:2.05,-0.06)node[right]{\footnotesize resonance}to(res);
				\end{pgfonlayer}
			\end{axis}
		\end{tikzpicture}
		\caption{Dependence of the complex energy eigenvalues on the scaling angle $\theta$ for $\prescript{8}{\phi}{\text{Be}}(4_1^+)$ using the $V_{\alpha\phi}^{3/2}(r)$ potential with an $\alpha$-particle rms matter radius of 1.84 fm. \label{fig:phiBe}}
	\end{figure}

Notably, when the $\alpha$-particle rms radius $R_\alpha$ is set to 1.84~fm, the $\prescript{8}{\phi}{\text{Be}}(4^+_1)$ state shifts from a bound state to a resonant state using the $V_{\alpha\phi}^{3/2}(r)$ potential, with its position in the complex plane shown in Fig.~\ref{fig:phiBe}. Whether $\prescript{8}{\phi}{\text{Be}}(4^+_1)$ is a resonant state or a bound state remains an open question, warranting further theoretical and experimental investigation. 
    
	\subsection{The $\prescript{8}{J/\psi}{\text{Be}}$ and $\prescript{8}{\eta_c}{\text{Be}}$ systems}
 
    As shown in Table~\ref{tab:energy spectra}, the $J/\psi$ meson can form a weakly bound state with an $\alpha$ particle using the $V_{\alpha J/\psi}^{3/2}$, $V_{\alpha J/\psi}^{1/2}$, and $V_{\alpha J/\psi}^{\text{ave}}$ potentials, inconsistent with the findings in Ref.~\cite{Etminan:2025yoq}. Additionally, there are also some differences in the calculation of $\prescript{8}{M}{\text{Be}}$ energy levels. 
    These discrepancies may stem from three factors. First, Ref.~\cite{Etminan:2025yoq} used the Woods-Saxon potential, which is shallower than the analytical expressions we used. Second, the $\alpha$-particle rms radius $R_\alpha$ used in Ref.~\cite{Etminan:2025yoq} is 1.47 fm. Third, even when adopting the same Woods–Saxon potential and parameters as Ref.~\cite{Etminan:2025yoq}, we are unable to reproduce the results obtained via the hyperspherical harmonics method in that reference using GEM. 

A Borromean state is a stable three-body bound state in which all two-body subsystems are unbound. Ref.~\cite{Etminan:2025yoq} proposed that $\prescript{8}{J/\psi}{\text{Be}}$ could be a Borromean nucleus, as they did not find bound states for the $\alpha$-$J/\psi$ two-body system using the $V_{\alpha J/\psi}^{3/2}$ and $V_{\alpha J/\psi}^{\text{ave}}$ potentials. However, our calculations show that $\alpha$ and $J/\psi$ do form two-body bound states using the $V_{\alpha J/\psi}^{3/2}$, $V_{\alpha J/\psi}^{1/2}$, and $V_{\alpha J/\psi}^{\text{ave}}$ potentials. Hence, $\prescript{8}{J/\psi}{\text{Be}}$ is not a Borromean nucleus. On the other hand, our results suggest that $\prescript{8}{\eta_c}{\text{Be}}$ may indeed be a Borromean nucleus, since none of its binary subsystems can form a two-body bound state. This conclusion is consistent with Ref.~\cite{Etminan:2025yoq}, although there are differences in the calculated binding energies. 

The $\Lambda$ hyperon can bind the $0^+_1$ and $2^+_1$ states of $^8$Be, while the $\phi$ meson can bind the $0^+_1$, $2^+_1$, and $4^+_1$ states. In contrast, as shown in Fig.~\ref{fig:energy spectra}, $J/\psi$ and $\eta_c$ can only bind the lowest $0^+_1$ state, forming a shallow bound state---indicating that the interaction between charmonium and the nucleus is relatively weak. As shown in Table~\ref{tab:energy spectra}, for the $0^+_1$ bound state, the $\alpha$–$\alpha$ distance $R_{\alpha\alpha}$ significantly decreases, reflecting the glue-like roles of charmonium. Unlike the $\phi$ meson case, adding charmonium increase  $R_{\alpha\alpha}$ for the $2^+_1$ and $4^+_1$ resonant states. In fact, the lower the energy $E$ of the resonant states, the larger $R_{\alpha\alpha}$ becomes, which is the opposite trend 
observed for the $0^+_1$ bound states. Moreover, compared with $\prescript{8}{\phi}{\text{Be}}$, the $\alpha$-$M$ distance $R_{\alpha M}$ become smaller in $\prescript{8}{J/\psi}{\text{Be}}$. This occurs because the $\alpha$-$J/\psi$ interaction is relatively weak, while the $\alpha$-$\alpha$ interaction dominates. 

\begin{figure*}[htbp]
		\begin{tikzpicture}
			\begin{axis}[
				title=$\prescript{8}{J/\psi}{\text{Be}}(2_1^+)$,
				xlabel=$E_r$ (MeV),
				xmin=0, xmax=5, 
				ylabel=$-\Gamma_r/2$ (MeV),
				legend entries={$\theta=15^\circ$,$\theta=20^\circ$},
				legend pos=south west,
				mark size=1pt,
				set layers,
				]
				\addplot+[only marks,mark=square*]coordinates{
					(0.195378,-0.0764754)
					(0.27532,-0.123001)
					(0.334428,-0.142389)
					(0.412148,-0.186568)
					(0.426573,-0.210124)
					(0.546158,-0.247867)
					(0.566883,-0.276243)
					(0.621404,-0.290213)
					(0.691819,-0.363217)
					(0.773892,-0.376897)
					(0.829312,-0.427686)
					(0.878131,-0.418319)
					(0.950183,-0.457007)
					(1.0416,-0.530971)
					(1.0996,-0.542953)
					(1.13784,-0.620709)
					(1.27546,-0.684842)
					(1.36521,-0.688486)
					(1.41156,-0.688785)
					(1.48227,-0.725064)
					(1.48547,-0.788189)
					(1.62929,-0.805357)
					(1.81614,-0.949223)
					(1.89166,-0.958016)
					(1.88349,-1.05137)
					(2.2238,-0.207167)
					(2.01952,-1.11416)
					(2.22835,-1.21594)
					(2.36282,-1.07787)
					(2.33749,-1.20811)
					(2.42785,-1.12406)
					(2.56567,-1.21214)
					(2.55709,-1.38066)
					(2.96206,-0.664825)
					(2.81672,-1.3698)
					(3.05794,-0.719076)
					(3.2469,-0.827555)
					(3.08618,-1.64088)
					(3.12586,-1.76889)
					(3.25172,-1.64083)
					(3.58876,-1.02503)
					(3.26267,-1.83163)
					(3.46714,-1.93081)
					(3.79152,-2.09189)
					(4.1828,-1.38053)
					(3.9357,-2.0988)
					(4.0993,-2.23004)
					(4.16558,-2.2459)
					(4.27016,-2.31986)
					(4.34069,-2.35876)
					(4.51859,-2.454)
					(4.89968,-2.69705)
					(5.25271,-2.02837)
				};
				
				\addplot+[only marks,mark=*]coordinates{
					(0.183459,-0.099623)
					(0.254025,-0.159401)
					(0.310661,-0.184871)
					(0.379707,-0.241767)
					(0.387886,-0.271406)
					(0.502868,-0.321181)
					(0.516458,-0.35711)
					(0.569812,-0.375674)
					(0.62254,-0.468241)
					(0.705328,-0.48748)
					(0.748589,-0.551786)
					(0.802506,-0.541658)
					(0.867349,-0.591926)
					(0.94235,-0.686064)
					(0.999466,-0.702304)
					(1.01708,-0.799264)
					(1.14339,-0.882389)
					(1.23652,-0.892474)
					(1.28679,-0.901291)
					(1.35006,-0.949527)
					(1.3344,-1.01634)
					(1.47954,-1.05464)
					(1.63456,-1.22724)
					(1.70918,-1.25038)
					(2.10813,-0.210419)
					(1.67657,-1.35291)
					(1.80175,-1.43431)
					(1.99183,-1.5662)
					(2.1322,-1.50711)
					(2.10464,-1.56547)
					(2.19119,-1.55819)
					(2.31171,-1.6595)
					(2.29085,-1.78108)
					(2.83577,-0.649315)
					(2.9189,-0.723019)
					(2.5292,-1.84904)
					(3.08751,-0.870107)
					(2.76434,-2.12839)
					(3.38687,-1.13779)
					(2.77544,-2.27528)
					(2.90902,-2.17941)
					(2.90136,-2.35681)
					(3.08818,-2.48522)
					(3.92181,-1.61566)
					(3.38163,-2.69372)
					(3.48093,-2.72136)
					(3.52168,-2.87719)
					(3.59189,-2.90509)
					(3.6904,-2.98934)
					(3.85648,-3.06646)
					(3.90611,-3.16481)
					(4.8567,-2.45194)
					(4.24334,-3.46819)
					(4.64817,-3.69222)
					(4.60627,-3.81178)
					(4.7328,-3.89432)
					(4.89369,-3.98414)
					(4.92058,-4.02056)
					(5.20985,-4.22545)
				};

				\draw[gray,thin](axis cs:0,0)--(axis cs:5,0);
				
				\begin{pgfonlayer}{axis foreground}
					\draw[semithick](axis cs:2.83577,-0.649315)--(axis cs:4.8567,-2.45194);
					\draw[-stealth](axis cs:4.1,-1)node[above,align=center,font=\footnotesize]{$\prescript{8}{}{\text{Be}}(2^+_1)+J/\psi$\\continuum}--(axis cs:3.7,-1.3);
					\draw[semithick](axis cs:0.183459,-0.099623)--(axis cs:4.9,-3.95);
					\draw[-stealth](axis cs:2.1,-2.4)node[below,align=center,font=\footnotesize]{$\alpha+\alpha+J/\psi$\\continuum}--(axis cs:2.5,-2.1);
					\node[circle,draw,inner sep=4](res)at(axis cs:2.17,-0.210419){};
					\draw[-stealth](axis cs:2.55,-0.1)node[right]{\footnotesize resonance}to(res);
				\end{pgfonlayer}
			\end{axis}
		\end{tikzpicture}
		\begin{tikzpicture}
			\begin{axis}[
				title=$\prescript{8}{J/\psi}{\text{Be}}(4_1^+)$,
				xlabel=$E_r$ (MeV),
				xmin=0, xmax=30, 
				ylabel=$-\Gamma_r/2$ (MeV),
				legend entries={$\theta=15^\circ$,$\theta=20^\circ$},
				legend pos=south west,
				mark size=1pt,
				set layers,
				]

				\addplot+[only marks,mark=square*]coordinates{
					(0.0773538,-0.0329108)
					(0.109805,-0.0473499)
					(0.156069,-0.068742)
					(0.156053,-0.0783467)
					(0.188472,-0.092753)
					(0.22565,-0.102262)
					(0.234766,-0.114253)
					(0.305195,-0.148486)
					(0.308463,-0.166342)
					(0.3332,-0.155911)
					(0.340899,-0.180775)
					(0.387418,-0.20244)
					(0.412669,-0.201915)
					(0.456994,-0.235875)
					(0.50217,-0.242662)
					(0.564912,-0.290002)
					(0.580139,-0.287334)
					(0.573255,-0.31922)
					(0.605716,-0.333684)
					(0.652325,-0.35542)
					(0.721898,-0.388873)
					(0.734323,-0.3766)
					(0.77075,-0.384003)
					(0.829435,-0.442346)
					(0.84647,-0.426549)
					(1.0002,-0.51529)
					(0.997282,-0.528487)
					(1.01935,-0.576773)
					(1.05183,-0.59126)
					(1.09848,-0.61304)
					(1.16835,-0.646751)
					(1.20166,-0.614603)
					(1.27277,-0.655639)
					(1.26878,-0.669103)
					(1.27523,-0.699689)
					(1.42563,-0.741546)
					(1.44356,-0.785593)
					(1.69295,-0.895293)
					(1.70861,-0.924234)
					(1.76446,-1.00696)
					(1.79695,-1.02146)
					(1.84364,-1.0433)
					(1.89876,-0.993962)
					(1.91367,-1.07716)
					(1.96876,-1.03118)
					(2.02092,-1.13039)
					(2.11339,-1.11313)
					(2.14242,-1.15564)
					(2.18773,-1.21508)
					(2.37878,-1.26556)
					(2.45418,-1.35365)
					(2.82781,-1.52118)
					(2.87733,-1.57858)
					(3.03424,-1.61854)
					(3.00646,-1.72403)
					(3.03896,-1.73854)
					(3.10116,-1.65271)
					(3.08566,-1.7604)
					(3.15582,-1.79441)
					(3.24137,-1.73073)
					(3.26345,-1.84799)
					(3.43071,-1.93288)
					(3.49584,-1.87469)
					(3.57698,-1.9556)
					(3.69417,-2.06925)
					(3.9419,-2.12978)
					(4.11939,-2.2934)
					(4.68764,-2.54544)
					(4.80083,-2.6594)
					(4.90504,-2.64565)
					(4.96898,-2.67631)
					(5.10317,-2.74915)
					(5.07605,-2.91891)
					(5.10855,-2.93343)
					(5.15527,-2.95529)
					(5.22548,-2.98939)
					(5.34974,-2.88705)
					(5.33343,-3.04331)
					(5.50153,-3.1289)
					(5.77922,-3.1271)
					(5.76538,-3.26514)
					(5.93643,-3.27178)
					(6.18514,-3.48554)
					(6.52346,-3.54813)
					(6.86789,-3.84857)
					(7.76417,-4.22241)
					(7.97659,-4.44826)
					(8.16246,-4.28186)
					(8.22151,-4.31158)
					(8.35166,-4.38307)
					(8.59107,-4.5175)
					(8.52517,-4.91026)
					(8.55768,-4.92476)
					(8.60438,-4.94666)
					(8.67464,-4.98077)
					(8.78275,-5.03492)
					(9.01051,-4.75547)
					(10.149,-0.91895)
					(8.95158,-5.12131)
					(9.21721,-5.25878)
					(9.72473,-5.16291)
					(9.6369,-5.47791)
					(9.84082,-5.42753)
					(11.6777,-1.55499)
					(10.3102,-5.83462)
					(11.7685,-1.60934)
					(11.9492,-1.71806)
					(12.2755,-1.91568)
					(10.9594,-5.86659)
					(12.8514,-2.26616)
					(11.4132,-6.4222)
					(13.8618,-2.88224)
					(12.9661,-6.99827)
					(13.2446,-7.40007)
					(14.1584,-7.40109)
					(14.1858,-7.52082)
					(15.6326,-3.95295)
					(14.3067,-7.58017)
					(14.5298,-7.70333)
					(14.2749,-8.22987)
					(14.3074,-8.24439)
					(14.3541,-8.26623)
					(14.4244,-8.30044)
					(14.5326,-8.35463)
					(14.918,-7.92366)
					(14.7018,-8.44155)
					(14.969,-8.58084)
					(15.5821,-8.30601)
					(15.3923,-8.80191)
					(16.0638,-9.154)
					(16.3993,-8.93089)
					(16.7977,-8.98323)
					(18.7039,-5.75666)
					(17.1507,-9.72871)
					(18.6705,-10.0481)
					(18.9614,-10.6684)
					(21.6378,-11.8088)
					(22.2686,-12.2128)
					(23.9421,-8.81935)
					(23.1019,-12.2014)
					(23.2295,-12.4181)
					(23.3361,-12.4741)
					(23.5431,-12.5884)
					(23.9081,-12.7955)
					(23.8634,-13.7658)
					(23.8959,-13.7803)
					(23.9426,-13.8022)
					(24.0129,-13.8363)
					(24.1211,-13.8907)
					(24.5366,-13.1576)
					(24.2906,-13.9778)
					(24.5589,-14.1182)
					(24.9857,-14.3431)
					(25.6019,-13.7791)
					(25.6637,-14.6974)
					(26.7439,-15.2589)
					(27.186,-14.8125)
					(27.8972,-15.1961)
					(28.5282,-16.1704)
					(30.6464,-16.6977)
				};
				
				\addplot+[only marks,mark=*]coordinates{
					(0.0718019,-0.0426879)
					(0.101751,-0.0613955)
					(0.144225,-0.0890871)
					(0.141416,-0.101099)
					(0.171342,-0.119766)
					(0.207704,-0.132411)
					(0.213802,-0.14758)
					(0.277946,-0.191806)
					(0.27623,-0.214223)
					(0.305278,-0.201666)
					(0.306164,-0.232922)
					(0.348812,-0.260948)
					(0.37553,-0.260804)
					(0.412319,-0.304172)
					(0.457866,-0.313566)
					(0.510111,-0.374039)
					(0.526977,-0.371021)
					(0.510453,-0.41076)
					(0.540401,-0.429496)
					(0.583122,-0.457609)
					(0.646611,-0.500874)
					(0.663297,-0.485833)
					(0.699371,-0.495684)
					(0.744249,-0.569918)
					(0.766841,-0.550548)
					(0.902851,-0.664685)
					(0.895819,-0.681024)
					(0.905044,-0.741863)
					(0.935005,-0.760626)
					(0.977752,-0.788798)
					(1.04147,-0.83236)
					(1.08586,-0.792872)
					(1.14974,-0.844818)
					(1.13999,-0.863381)
					(1.13857,-0.900764)
					(1.2849,-0.95629)
					(1.29083,-1.01163)
					(1.5217,-1.15422)
					(1.52946,-1.19039)
					(1.56414,-1.29491)
					(1.5941,-1.31369)
					(1.63687,-1.34191)
					(1.70912,-1.28146)
					(1.70072,-1.38569)
					(1.77232,-1.32964)
					(1.79813,-1.45442)
					(1.90097,-1.43511)
					(1.91884,-1.48887)
					(1.94913,-1.5638)
					(2.13591,-1.63121)
					(2.18946,-1.74261)
					(2.53492,-1.96095)
					(2.56943,-2.03258)
					(2.72242,-2.08667)
					(2.66275,-2.21675)
					(2.69272,-2.23554)
					(2.78341,-2.13102)
					(2.7355,-2.2638)
					(2.7994,-2.30773)
					(2.90844,-2.23148)
					(2.89714,-2.37695)
					(3.04856,-2.48656)
					(3.1347,-2.41702)
					(3.19691,-2.51931)
					(3.28634,-2.6626)
					(3.53028,-2.74566)
					(3.66923,-2.95187)
					(4.19632,-3.28423)
					(4.27993,-3.42419)
					(4.38824,-3.42547)
					(4.44719,-3.46561)
					(4.5671,-3.5597)
					(4.49341,-3.75286)
					(4.52338,-3.77165)
					(4.56616,-3.79992)
					(4.63012,-3.84396)
					(4.78631,-3.7379)
					(4.72802,-3.91357)
					(4.88016,-4.02412)
					(5.16877,-4.04873)
					(5.11851,-4.20004)
					(5.29943,-4.21713)
					(5.4965,-4.4846)
					(5.82921,-4.59321)
					(6.1105,-4.95331)
					(6.93503,-5.47011)
					(7.10104,-5.72873)
					(7.29242,-5.71024)
					(7.34618,-5.74496)
					(7.45949,-5.83564)
					(7.66955,-6.00782)
					(7.54434,-6.31289)
					(7.57431,-6.33168)
					(7.61709,-6.35997)
					(7.68106,-6.40405)
					(7.77909,-6.47391)
					(10.1481,-0.922268)
					(8.03916,-6.31195)
					(7.93163,-6.58541)
					(8.17147,-6.76303)
					(8.66943,-6.8244)
					(8.55025,-7.04614)
					(8.7797,-7.02479)
					(11.6643,-1.58421)
					(9.15593,-7.50707)
					(11.7434,-1.6544)
					(11.9006,-1.79494)
					(12.1839,-2.05059)
					(9.76811,-7.7283)
					(12.6845,-2.50403)
					(10.1475,-8.26849)
					(13.5652,-3.29976)
					(11.5612,-9.15529)
					(11.7846,-9.56077)
					(15.1167,-4.67608)
					(12.5969,-9.83392)
					(12.5465,-9.9334)
					(12.6629,-10.0081)
					(12.8673,-10.1681)
					(12.6303,-10.5805)
					(12.6602,-10.5993)
					(12.703,-10.6276)
					(12.767,-10.6717)
					(12.865,-10.7417)
					(13.2188,-10.4504)
					(13.0179,-10.8538)
					(13.2588,-11.0335)
					(13.8202,-10.9334)
					(13.6406,-11.3195)
					(14.2467,-11.7752)
					(14.596,-11.6492)
					(14.9108,-11.8413)
					(17.8094,-7.00376)
					(15.2258,-12.5211)
					(16.6082,-13.1499)
					(16.8573,-13.7674)
					(19.3162,-15.2604)
					(22.4124,-10.9593)
					(19.7988,-15.9859)
					(20.8014,-15.6593)
					(20.9829,-15.9503)
					(21.0742,-16.0232)
					(21.2535,-16.1693)
					(21.5713,-16.4318)
					(21.1118,-17.6973)
					(21.1418,-17.7161)
					(21.1845,-17.7444)
					(21.2486,-17.7885)
					(22.1197,-16.89)
					(21.3466,-17.8586)
					(21.4995,-17.9711)
					(21.7412,-18.1522)
					(22.1256,-18.4428)
					(23.0494,-17.6767)
					(22.7382,-18.902)
					(23.7156,-19.6335)
					(24.4465,-19.0413)
					(24.8791,-19.6745)
					(25.3243,-20.8486)
					(27.4534,-21.4222)
					(30.2406,-17.6195)
					(28.1263,-22.9512)
					(31.3288,-25.2984)
				};
				
				\draw[gray,thin](axis cs:0,0)--(axis cs:30,0);
				
				\begin{pgfonlayer}{axis foreground}
					\draw[semithick](axis cs:11.6643,-1.58421)--(axis cs:22.4124,-10.9593);
					\draw[-stealth](axis cs:19,-4.5)node[right,align=center,font=\footnotesize]{$\prescript{8}{}{\text{Be}}(4^+_1)+J/\psi$\\continuum}--(axis cs:17,-6);
					\draw[semithick](axis cs:0,0)--(axis cs:28,-22.7);
					\draw[-stealth](axis cs:13,-14.5)node[below,align=center,font=\footnotesize]{$\alpha+\alpha+J/\psi$\\continuum}--(axis cs:15,-13);
					\node[circle,draw,inner sep=2](res)at(axis cs:10.1481,-0.922268){};
					\draw[-stealth](axis cs:12,-0.6)node[right]{\footnotesize resonance}to(res);
				\end{pgfonlayer}
			\end{axis}
		\end{tikzpicture}
		\caption{Dependence of the complex energy eigenvalues on the scaling angle $\theta$ for $2_1^+$ and $4^+_1$ states of $\prescript{8}{J/\psi}{\text{Be}}$ using the $V_{\alpha J/\psi}^{3/2}(r)$ potential with an $\alpha$-particle rms matter radius of 1.70 fm. \label{fig:JpsiBe}}
	\end{figure*}
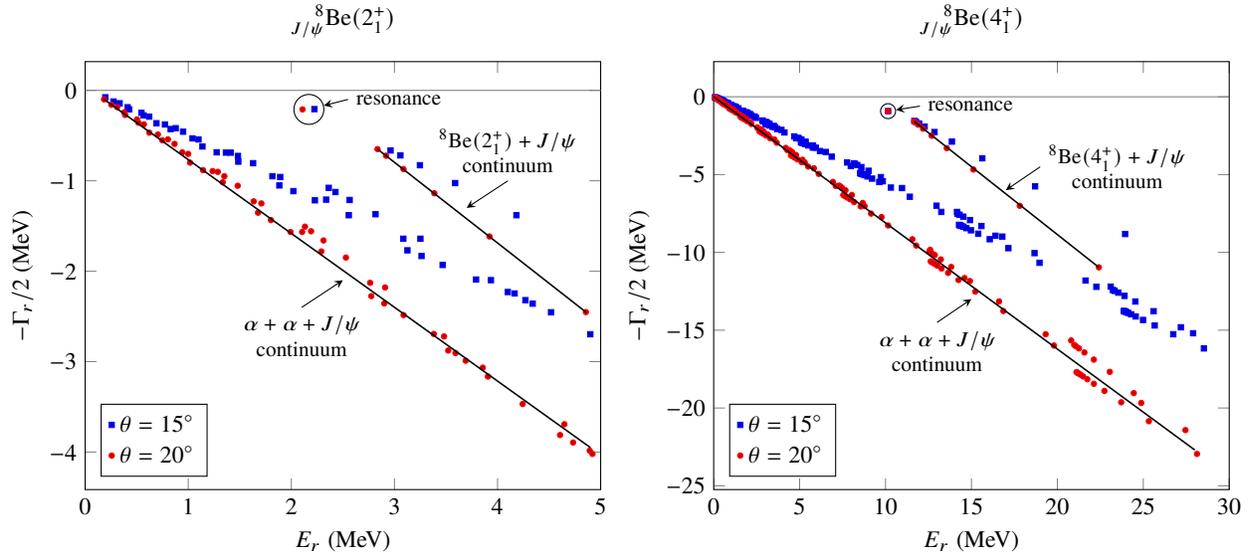

The $2^+_1$ and $4^+_1$ states of $\prescript{8}{J/\psi}{\text{Be}}$ remain resonant; their positions in the complex plane are shown in Fig.~\ref{fig:JpsiBe} with an $\alpha$-particle rms matter radius of 1.70 fm. The situation is similar for other systems, and is therefore not detailed here. It can be seen that the $2^+_1$ resonance lies slightly below—and close to—the $\prescript{8}{}{\text{Be}}(2^+_1)+J/\psi$ threshold, while the $4^+_1$ resonance lies slightly below—and also close to—the $\prescript{8}{}{\text{Be}}(4^+)+J/\psi$ threshold.

	\section{SUMMARY}\label{sec4}
    
    Based on the recently obtained HAL QCD potentials for interactions between nucleons ($N$) and the $\phi$, $J/\psi$, and $\eta_c$ mesons, this paper systematically investigates the structure of the exotic nuclei $\prescript{8}{\phi}{\text{Be}}$, $\prescript{8}{J/\psi}{\text{Be}}$, and $\prescript{8}{\eta_c}{\text{Be}}$. These systems are modeled as three-body clusters of two alpha particles ($\alpha$) and a meson ($M$). The study employs the GEM to solve the Schr\"odinger equation, allowing for a consistent treatment of both bound and resonant states. Resonant states are specifically identified using the CSM. To construct the effective interaction between an alpha particle and a meson ($\alpha$-$M$), a folding potential is generated from the underlying nucleon-meson HAL QCD potentials and the unanalytical potential is fitted to a Woods-Saxon form for computational implementation.

Our findings reveal different roles played by the different mesons in modifying the structure of the $^8$Be core (treated as an $\alpha$-$\alpha$ dimer). The $\phi$ meson exhibits a "glue-like" effect. Its strong attraction not only binds the excited resonant states of $^8$Be—specifically the $0^+_1$, $2^+_1$, and $4^+_1$ states—into stable, bound states within $\prescript{8}{\phi}{\text{Be}}$, but also induces a significant contraction of the distance between the two alpha particles. This effect is analogous to phenomena previously observed with the $\Lambda$ hyperon in hypernuclei.

In contrast, the interactions of the charmonium states $J/\psi$ and $\eta_c$ with nuclear matter are considerably weaker, as suggested by the OZI rule and the underlying HAL QCD potentials. Consequently, these mesons form only shallow bound states with the ground ($0^+_1$) state of $^8$Be. Notably, instead of contracting the core, the addition of a $J/\psi$ or $\eta_c$ meson slightly increases the separation between the alpha clusters. A result is the prediction of weakly bound states in the $\alpha$-$J/\psi$ two-body subsystem using the $V_{\alpha J/\psi}^{3/2}$, $V_{\alpha J/\psi}^{1/2}$, and $V_{\alpha J/\psi}^{\text{ave}}$ potentials. This finding implies that $\prescript{8}{J/\psi}{\text{Be}}$ may not be a Borromean system (where binding occurs only in the three-body system but not in any two-body subsystem), a distinction not highlighted in prior studies~\cite{Etminan:2025yoq}.

We note that Ref.~\cite{Metag:2017yuh,Krein:2017usp} discusses the search for $\phi$-nucleus and charm-nucleus bound states. At the time, however, the $\phi$-nucleus attraction was believed to be rather weak—even the $\phi{}^{11}\text{B}$ potential was considered too shallow to support a $\phi$–nucleus bound state. Nevertheless, an interesting experiment (E29)~\cite{J-PARC:E29} using an antiproton beam has been proposed at J-PARC to investigate $\phi$-mesic states.
As for the $\prescript{8}{J/\psi}{\text{Be}}$ and $\prescript{8}{\eta_c}{\text{Be}}$ systems, the main experimental challenge lies in the large momentum transfer involved in charm production. The 12 GeV upgrade at JLab has significantly enhanced its capabilities for nuclear physics research, consequently enabling new experiments such as a search for the predicted states. These predicted exotic nuclei have yet to be explored experimentally. With the progress of experiments, we have reason to believe that these predicted exotic nuclei could become accessible in future experiments, forming an issue that awaits exploration.

\vfil

\begin{acknowledgments}

This work is also supported by the National Natural Science Foundation of China under Grant Nos. 12335001 and 12247101, the ‘111 Center’ under Grant No. B20063, the Natural Science Foundation of Gansu Province (No. 22JR5RA389, No. 25JRRA799), the fundamental Research Funds for the Central Universities, the project for top-notch innovative talents of Gansu province, and Lanzhou City High-Level Talent Funding.
 
\end{acknowledgments}


	\bibliography{References}
	
\end{document}